\documentclass[twocolumn]{aastex631}
\usepackage[utf8]{inputenc}
\usepackage[T1]{fontenc}
\usepackage{hyperref}
\usepackage{amsmath}
\usepackage{newunicodechar}
\usepackage{threeparttable}

\newunicodechar{́}{\'{}}

\received{September 3, 2025}
\revised{December 1, 2025}
\accepted{December 2, 2025}
\submitjournal{ApJ}
\begin{document}

\title{Astrometric and Spectroscopic Analysis of IC 2714: \\ An Open Cluster Hosting a Lithium-Rich Giant}

\author[0009-0008-0404-4031]{T. Flaulhabe}
\affiliation{Observatório Nacional, Rua General José Cristino 77, CEP 20921-400, São Cristóvão, Rio de Janeiro, RJ, Brazil}

\author[0000-0002-8504-6248]{N. Holanda}
\affiliation{Observatório Nacional, Rua General José Cristino 77, CEP 20921-400, São Cristóvão, Rio de Janeiro, RJ, Brazil}

\author[0000-0001-7672-154X]{G. Tautvaišienė}
\affiliation{Vilnius University, Faculty of Physics, Institute of Theoretical Physics and Astronomy, Sauletekio av. 3, 10257 Vilnius, Lithuania}

\author[0000-0003-0439-2331]{O. J. Katime Santrich}
\affiliation{Universidade Estadual de Santa Cruz, UESC, Rodovia Jorge Amado km 16, Ilh\'eus 45662000, Bahia, Brazil}

\author[0000-0002-2569-4032]{F. F. S. Maia}
\affiliation{Instituto de Física, Universidade Federal do Rio de Janeiro, UFRJ, 21941-972, Rio de Janeiro, RJ, Brazil}
\affiliation{Observatório Nacional, Rua General José Cristino 77, CEP 20921-400, São Cristóvão, Rio de Janeiro, RJ, Brazil}

\author[0000-0002-7552-3063]{B. P. L. Ferreira}
\affiliation{Departamento de Astronomia, Instituto de Astronomia, Geofísica e Ciências Atmosféricas, Universidade de São Paulo, Rua do Matão 1226, Cidade Universitária, São Paulo 05508-090, Brazil}
\affiliation{Departamento de Física, Instituto de Ciências Exatas, Universidade Federal de Minas Gerais, UFMG, Av. Antônio Carlos 6627, Belo Horizonte 31270-901, Brazil}

\author[0000-0002-5150-5282]{W. J. B. Corradi}
\affiliation{Laboratório Nacional de Astrofísica, R. dos Estados Unidos, 154 – Nações, Itajubá, MG 37504-364, Brazil}
\affiliation{Departamento de Física, Instituto de Ciências Exatas, Universidade Federal de Minas Gerais, UFMG, Av. Antônio Carlos 6627, Belo Horizonte 31270-901, Brazil}

\author{C. B. Pereira}
\affiliation{Observatório Nacional, Rua General José Cristino 77, CEP 20921-400, São Cristóvão, Rio de Janeiro, RJ, Brazil}

\author[0000-0003-1757-6666]{M. Carlos}
\affiliation{Observatório Nacional, Rua General José Cristino 77, CEP 20921-400, São Cristóvão, Rio de Janeiro, RJ, Brazil}

\author[0000-0001-9205-2307]{S. Daflon}
\affiliation{Observatório Nacional, Rua General José Cristino 77, CEP 20921-400, São Cristóvão, Rio de Janeiro, RJ, Brazil}

\correspondingauthor{Thiago Flaulhabe}
\email{thiagogomes@on.br}

\begin{abstract}

Open clusters serve as laboratories to study and evaluate stellar evolution and Galactic chemical evolution models. Chemical peculiarities, such as lithium-rich giants, are rarely observed in these stellar systems. This work focuses on eight red giants (\#005, \#028, \#034, \#053, \#087, \#121, \#126, and \#190) previously reported as members of the Galactic cluster IC\,2714. We conducted a detailed investigation using high-resolution spectroscopy, supplemented with data from the Gaia\,DR3 catalog. Besides deriving the cluster's fundamental parameters, we provide the most thorough chemical characterization of IC\,2714 to date, reporting the abundance of 23 species, including light elements (Li, C, N, O), odd-Z elements (Na, Al), $\alpha$-elements (Si, Ca, Ti, Mg), iron-peak elements (Sc, Cr, Ni), \textit{s}-process-dominated elements (Y, Zr, Ba, La, Ce, Nd) and \textit{r}-process elements (Sm, Eu). We also present the carbon isotopic ratios $^{12}$C/$^{13}$C for the first time for seven stars. One particular star (\#087) exhibits a high lithium abundance ($\log\,\varepsilon$(Li)$_{\rm NLTE}$\,=\,$+$1.54\,dex) and a slightly higher projected rotational velocity ($v\,\sin\,i$\,=\,6.7\,\text{km\,s$^{-1}$}). Our results suggest that the analyzed stars are in the core-helium-burning phase of evolution, where the most lithium-rich giants are found. Combining astrometric probabilities and chemical abundances, we conclude that two giants (\#028 and \#034) might not be cluster members.

\end{abstract}


\section{Introduction} \label{sec:intro}

Open Clusters (OCs) are characterized by groups of gravitationally bound stars sharing similar distances, ages and initial chemical compositions. In this sense, they are invaluable tools for studying stellar and Galactic chemical evolution \citep[e.g.,][]{friel1995, lada2003, netopil2016, magrini2009}. Predominantly distributed across the Galactic disc, OCs provide the ideal conditions for testing stellar evolution models and to trace the chemical enrichment of the Galaxy.

In the last decade, the large datasets provided by the Gaia mission \citep{gaia2016} allowed a significant development of OCs studies in the Galaxy. Using Gaia's astrometric and photometric data, several works reported the discovery of a large number of clusters, including the application of unsupervised machine learning techniques to this intent \citep[e.g.][]{cantat2018b, ferreira2020, ferreira2019, Cantat2020, jaehnig2021, ferreira2021, hunt2023, cantat2019, castro2018, castro2019, castro2020, castro2022, liu2019}. 
Moreover, high-resolution spectroscopy proves to be an excellent tool for conducting a thorough chemical analysis of the cluster member stars. The abundances analysis can provide valuable insights of OC properties and also guide studies in the context of chemical gradients in the Galaxy \citep{netopil2016, magrini2009, magrini2017, magrini2023}.

This paper presents a comprehensive astrometric and spectroscopic analysis of the Galactic open cluster IC\,2714. Previous studies, such as \citet{Cantat2020} and \citet{hunt2023}, estimate its age at $0.44-0.52$ Gyr, with A$_\text{V} = 0.95$. Our analysis updates these results and provides detailed chemical abundances for 23 chemical species in eight red giants previously classified as members of IC\,2714.

To date, chemical analysis using high-resolution spectroscopy of the stars in our sample were limited to iron and lithium abundances \citep[e.g.][]{santos2009, delgadomena2016, tsantaki2023}. An exception is found in \citet{smiljanic2009}, where an analysis of several chemical species is presented for the star \#005. Furthermore, the star \#034 was studied by \citet{ramos2024}, which conducted an extensive chemical study and membership analysis of 17 single-lined spectroscopic binary stars in 15 OCs. For the star \#028, however, atmospheric parameters and chemical abundances derived via high-resolution spectroscopy are here presented for the first time.

A point of interest is the report of three lithium-rich giants within IC\,2714 \citep{delgadomena2016}, which comprise a very rare class of stars that represent only 1 to 2\% of the G$-$K giants in the Galaxy \citep[e.g.][]{magrini2021,cai2023, gao2022, smiljanic2018, wallerstein1982}. The $^7$Li atom is lithium's most abundant isotope and is easily destroyed by the first dredge-up (FDU) before a star enters the Red Giant Branch (RGB). In this sense, standard evolution models do not predict the existence of lithium-rich giants. In particular, standard models predict a photospheric abundance of $\log\,\varepsilon$(Li)\,$<$\,1.50\,dex after the FDU \citep{charbonnel2000}, and abundances above this limit are traditionally associated with chemically peculiar stars. Several mechanisms are presented in literature to explain this enrichment \citep[e.g.][]{siess1999b, alexander1967, zhang2020, gratton1989, kirby2016, fekel1993, sackmann1999, holanda2020a, schwab2020, mori2020, li2021, mallick2025}. 

It is currently understood that, due to the Hot Bottom Burning process (HBB), lithium atoms can be synthesized via the Cameron-Fowler mechanism \citep{cameron1971} in intermediate-mass stars during the Thermal Pulses phase in Asymptotic Giant Branch (TP-AGB). However, lithium enrichment in earlier evolutionary stages is still poorly understood \citep[e.g.][]{deepak2021, holanda2020a, holanda2020b, magrini2021}. For example, thermohaline mixing is suited to explain the low $^{12}$C/$^{13}$C isotopic ratios found in red giants, but it is unable to explain lithium enrichment before TP-AGB \citep{charbonnel2000, lagarde2019}. In this sense, \citet{holanda2020a} reported a Li-rich in the Early-AGB phase and suggested the Cool Bottom Process (CBP) as a possible mechanism to explain its enrichment. The CBP is mainly responsible for $^3$He destruction and  $^7$Li production, and could also explain the carbon isotopic ratios observed in RGB stars. However, as pointed out by the authors, the efficiency of this process depends on complex factors, such as mixing speeds, geometry and episodicity, and is not related to any physical explanation within stellar evolution models.

On the other hand, \citet{siess1999a,siess1999b} proposed a scenario in which the accretion of planets or of a substellar companion could be responsible for a lithium enrichment in the stellar photosphere, also increasing its rotational velocity. As proposed by the authors, this scenario could be corroborated by a photospheric overabundance of Be and B. However, \citet{takeda2017} found no Be enrichment among a large sample of Li-rich and non-rich giants. Also, \citet{drake2018} did not found any relevant B enrichment in four Li-rich giants. To explain these observations, \citet{holanda2020a} proposed that continuous circulation processes could deplete Be and B abundances in CBP-enriched giants.

To explain the existence of ``super Li-rich giants'', with lithium abundances greater than meteoritic value \citep[3.26;][]{asplund2009}, \citet{zhang2020} present a merger scenario between a RGB and a He-white dwarf companion. Even though this scenario conforms to the stellar parameters, carbon isotopic ratios and lithium abundances found in most Li-rich giants \citep{holanda2024a}, this model does not take rotational velocities in consideration and it can only explain lithium enrichment in core-helium-burning red giants, known as red clump (RC) stars.

Also, literature reports a class of chemically peculiar stars that are commonly associated with Li-rich giants, known as Weak G-band (WGB) stars. These stars are characterized by weak CH molecular lines in Fraunhofer's G-band, indicative of low $^{12}$C abundances \citep{rao1978}, and present high N abundances. Notably, several works observe an incidence of about 50\% of Li-rich stars between WGBs \citep[e.g.][]{palacios2012, palacios2016, holanda2024b, maben2023b}. However, when compared with regular Li-rich giants, WGBs present significantly lower $^{12}$C/$^{13}$C ratios, and have essentially low rotational velocities \citep{holanda2024a}. The formation mechanism of such stars is still debated, and several scenarios are studied, including merger events such as the proposed by \citet{zhang2020}.

Given the above mentioned scenarios, it is clear that lithium enrichment in giants is a debated theme in current literature, still not fully understood. Particularly, the existence of Li-rich red giants still lacks adequate explanations within current stellar evolution models. The analysis presented in this work is contributing in studying the nature of these objects.

\par We summarize our work as follows: in Section\,\ref{sec:obs} we describe the data used in this work, as well as the membership analysis; Section \ref{sec:meth} describes our adopted methodology; our results are presented together with a comparison with literature data in Section\,\ref{sec:discuss}; and finally, a summary and considerations are delineated in Section\,\ref{sec:conc}.

\section{Astrometric and Spectroscopic Data}\label{sec:obs}

\begin{figure*}
    \centering
    \includegraphics[scale=0.5]{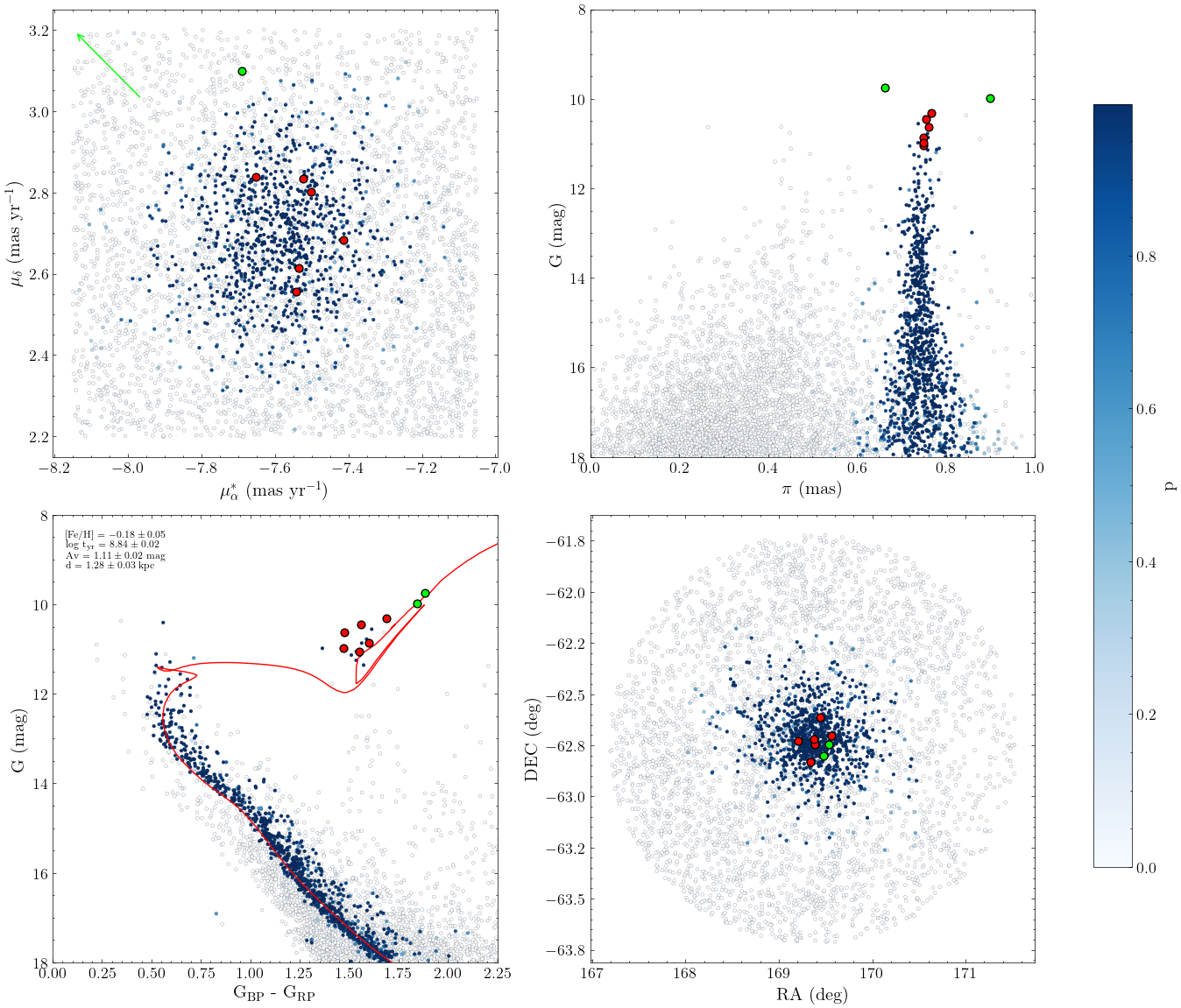}
    \caption{Proper-motion distribution of the extracted Gaia\,DR3 data around the center of IC\,2714 (top left), with the color scale indicating probabilities computed using pyUPMASK; the star \#034 is found outside the plot limits, as indicated by the green arrow. The top right panel shows the parallax versus G magnitude distribution. A color–magnitude diagram (bottom left) is presented, with an isochrone fit using {PARSEC} isochrones \citep{bressan2012}; the fit was performed using the SIESTA code, considering stars with p\,$\geq$\,0.7. The bottom right panel displays the spatial distribution of stars around the center of the cluster. The red giants of our sample classified as members are shown in red circles, while the non-member stars \#028 and \#034 are shown in green circles. In the top left plot, the star \#034 lies outside the plot limits, as indicated by the green arrow.}
    \label{fig:4plots}
\end{figure*}

In this work, we initially obtained astrometric and photometric data for all the stars contained in a 1° radius region centered on IC\,2714 from the Gaia DR3 catalogue \citep{gaia2023}. Quality filters (which will be described later) were then applied as part of the decontamination procedure. V and K-band photometry were obtained from \citet{mermilliod2008} and \citet{twomass2003}, respectively.

The eight giants in our sample were previously reported as cluster members based on their radial velocity (RV) by \citet{mermilliod2008}. We obtained high-resolution spectra of the stars during two missions at the La Silla Paranal Observatory in Chile, in 2009. The observations were conducted under collaborative agreements involving the Observat{\' o}rio Nacional in Brazil, the European Southern Observatory (ESO), and the Max-Planck-Institut für Astronomie (MPG).

\par We used the Fiber-fed Extended-Range Optical Spectrograph \citep[{\sc feros};][]{kaufer1999} attached to the 2.2\,m MPG/ESO telescope. \textsc{feros} offers a wavelength coverage ranging from 3700 to 9000\,\AA\, and a power resolution of \textit{R}\,$\approx$\,48,000. The spectral reduction was conducted with the {\sc feros} Data Reduction System\footnote{{\sc feros} Data Reduction system is described and available at \url{www.eso.org/sci/facilities/lasilla/instruments/feros/tools/DRS.html}}.

\par Table\,\ref{tab:general} provides general information about our sample. The eleventh column presents the signal-to-noise ratio (SNR) for each spectrum, measured around 6000\,\AA. The exposure times were estimated to ensure SNR values exceeding 100. The table also includes RV measurements from \citet{mermilliod2008}, who used these results to classify spectroscopic binary (SB) stars and candidates (SB?).  This classification is shown in the last column. Our RV measurements were obtained using {\sc ispec} \citep{cuaresma2014} with the template-matching fitting method. 

\begin{table*}
\centering
\caption{General information about the sample. The stars with p > 0.700 are here considered members, while the remaining (\#028 and \#034) are considered non-members. References: \citet{gaia2023}$^{a}$, \citet{mermilliod2008}$^b$.}
\label{tab:general}
\begin{tabular} {l c c c c c c c c c c c c c}
\hline
Star & $\mu^*_\alpha \text{ }^a $ & $\mu_\delta \text{ }^a$ & $G^a$ & $(G_{\rm BP}-G_{\rm RP})^a$ &  \texttt{ruwe}$^a$ & RV$^a$ & RV$^b$ & RV & p & SNR & Note$^b$ \\
     ~  &mas yr$^{-1}$    & mas yr$^{-1}$& mag & mag                     & ~        & km s$^{-1}$  & km s$^{-1}$  & km s$^{-1}$ & ~ & ~ & ~ \\
\hline                                                              
\#005 & $-$\,7.502  & 2.801 & 10.623 & 1.478 & 0.95 & $-$\,14.32 & $-$\,14.53 & $-$\,13.45\,$\pm$\,0.54 & 0.995 & 127 & - \\
\#028$^\star$ & $-$\,7.690  & 3.099 & 9.750  & 1.884 & 1.08 & $-$\,10.77 & $-$\,13.32 & $-$\,12.07\,$\pm$\,0.82 & 0.093 & 140 & SB \\
\#034$^\star$ & $-$\,11.509 & 7.622 & 9.985  & 1.842 & 0.88 & $-$\,17.46 & $-$\,17.76 & $-$\,17.69\,$\pm$\,0.59 & -     & 152 & SB?\\
\#053 & $-$\,7.651  & 2.837 & 11.050 & 1.554 & 0.88 & $-$\,13.09 & $-$\,13.37 & $-$\,13.35\,$\pm$\,0.66 & 0.994 & 135 & - \\
\#087 & $-$\,7.415  & 2.684 & 10.987 & 1.472 & 0.75 & $-$\,12.67 & $-$\,13.23 & $-$\,13.12\,$\pm$\,0.70 & 0.993 & 150 & - \\
\#121 & $-$\,7.522  & 2.834 & 10.308 & 1.688 & 0.96 & $-$\,13.11 & $-$\,13.37 & $-$\,13.45\,$\pm$\,0.54 & 0.995 & 161 & - \\
\#126 &$-$\,7.534   & 2.614 & 10.451 & 1.559 & 0.96 & $-$\,14.05 & $-$\,14.42 & $-$\,14.10\,$\pm$\,0.65 & 0.995 & 125 & - \\
\#190 &$-$\,7.543   & 2.556 & 10.666 & 1.602 & 0.96 & $-$\,13.09 & $-$\,13.60 & $-$\,13.69\,$\pm$\,0.63 & 0.993 & 120 & - \\
\hline
\end{tabular}
    \begin{tablenotes}
      \small
      \item Note: non-member stars are marked with $\star$.
    \end{tablenotes}
\end{table*}

\section{Methods}\label{sec:meth}

\subsection{Membership Analysis}

\par We implemented the Python version of Unsupervised Photometric Membership Assignment in Stellar Clusters method \citep[pyUPMASK;][]{krone2014,pera2021}\footnote{pyUPMASK is available at \url{https://github.com/msolpera/pyUPMASK}.} to perform a membership analysis for IC\,2714. This method is based on an iterative process that has already been proved to be effective in separating cluster and field star populations. PyUPMASK uses astrometric data, Principal Component Analysis (PCA) and a clustering algorithm to assign membership probabilities to the stars in the sample. In using pyUPMASK, we have adopted 10 outer loop runs, 10 stars per statistical cluster, and 3 PCA dimensions. Besides that, we have adopted the K-Means as clustering algorithm because its effectiveness has already been proven in analyzing bright OCs in the Galaxy \citep[e.g.][]{cantat2018, holanda2022, yontan2019}. The input sample was obtained by applying certain quality filters to the stars. First, we have discarded stars with negative Gaia parallaxes to ensure the quality of the data. Also, astrometric parallaxes were corrected following \citet{lindegren2021}. Then, considering that Gaia astrometric uncertainties become significant beyond G\,$=\,18$\,mag, we have considered only stars brighter than this limit. 
Besides, we excluded stars with \texttt{ruwe} $> 1.4$. The \texttt{ruwe} is a reliable statistical indicator provided by Gaia that allows to evaluate the quality of the astrometric solution, and values smaller than about 1.4 indicate good astrometric solutions. Values higher than this threshold are often associated with unresolved binaries \citep{lindegren2018}. Finally, in order to guarantee more sensitivity to cluster density and exclude possible contamination, we have constrained the proper motions space to $-8.15 \leq \mu^*_\alpha \leq -7.05$ mas yr$^{-1}$ and $2.20 \leq \mu_\delta \leq 3.20$ mas yr$^{-1}$. The final input astrometric sample used in pyUPMASK included 4498 stars. We considered as cluster members those stars with assigned astrometric membership p\,$\geq$\,0.700, and under this criteria our final sample contains 898 members. This probability threshold was found to provide the sample that best delineates the cluster main sequence, as can be seen in Figure \ref{fig:4plots}, while also reducing the number of outliers. The obtained completeness and purity for this sample were found to be greater than 85\% when compared to the member list provided by \citet{Cantat2020}. These scores show that the obtained sample is balanced, encompassing most of the known cluster members while presenting few contaminants.
Furthermore, the number of our member stars agrees with those from the literature, as will be discussed in \ref{sec:membership_results}.

\par Figure \ref{fig:4plots} shows our membership analysis results: at the top left, we present a bi-dimensional description of the stars in the astrometric input sample, centered at ($\mu^*_\alpha$, $\mu_\delta$) = ($-$7.59, $+$2.69) mas yr$^{-1}$. The stars are color-coded by its membership probability assigned by pyUPMASK. At the top right, a distribution of parallaxes around 0.744\,mas is presented. We can see how the cluster members are distributed around the average parallax of the cluster. In this figure, the six giants of our sample classified as members are shown in red, while the two giants with p\,$<$\,0.700 (\#028 and \#034) are shown in green. In both distributions, we can see how the member giants of our sample are located near the cluster's overdensities, while the non-members present discrepant proper motions and parallaxes: particularly, the star \#034 is found outside the limits of the proper motions distribution plot as indicated by the green arrow (top left). At the bottom left, we show the Color-Magnitude Diagram (CMD) for IC\,2714 with the isochrone fitting results, including metallicity, age, reddening and distance of the cluster. Finally, at the bottom right, the sky chart shows that all giants are located inside the cluster central region.

\par We have also conducted an isochrone fitting considering the member stars. For this purpose we used {\sc siesta}\footnote{{\sc siesta} is available at \url{https://github.com/Bereira/SIESTA}.} \citep{ferreira2024}, which allows us to conduct an isochrone fitting in a semi-automatic way, enabling stellar populations characterization from comparing the distribution of cluster members with synthetic populations generated from an isochrone grid. This algorithm had already been extensively applied to OCs in the context of the Magellanic Clouds \citep[e.g.][]{saroon2025} and here we present its first application to a Galactic OC. 

{\sc siesta} uses a Bayesian approach to determine the best-fit parameters that create the synthetic population that is the most representative of the observed data. This algorithm employs a Markov Chain Monte Carlo (MCMC) analysis, in which we adopted gaussian priors for the metallicity and distance. For the former, we adopted the average of the spectroscopic metallicities of the member giants in our sample. The distance prior was evaluated using the simple inverse parallax of the mean cluster members parallaxes, which yielded a result of $1.3 \pm 0.1$ kpc. This parameter is not expected to have a significant impact on the MCMC posterior distribution. Also, the quality filters applied to Gaia\,DR3 data assured a sample with fractional parallax errors ($f = \sigma_\pi/\pi$) less than 20\%. Since the inverse parallax approach becomes extremely noisy particularly for errors larger than about this limit \citep{bailer-jones2015}, we considered this estimate as acceptable in this case. For extinction and age, we used non-informative (e.g. uniform) priors adopting literature results for IC\,2714. In this work, we adopted {\sc parsec} isochrones \citep{bressan2012}. Using the best-fit isochrone, the cluster turn-off mass was estimated from the turning point in the mass vs. ($G_\text{BP}$$-$$G_\text{RP}$) relation, and the uncertainty was adopted as twice the isochrone mass bin.

\par In Table \ref{tab:isocfit} we present the main cluster parameters obtained via membership analysis and isochrone fitting. These values were obtained considering the mean values of the cluster members. Moreover, individual astrometric probabilities (p) of the stars in our sample are presented in Table\,\ref{tab:general}.

\begin{table}
\caption{Main cluster parameters. The values obtained in this work were calculated considering the mean values of the cluster members and assuming 8.34\,$\pm$\,0.16 kpc for the solar Galactocentric distance \citep{reid2014}. References: CG20: \citet{Cantat2020}; TS23: \citet{tsantaki2023}; H23: \citet{hunt2023}; Q24 DG and Q24 ML: \citet{qiu2024}, DG algorithm and maximum likelihood method, respectively.}
\label{tab:isocfit}
\begin{tabular} {l c c}
\hline
RA (deg)                        & \,\,169.383\,$\pm$\,0.327     & This Work \\
DEC (deg)                       & $-$62.715\,$\pm$ 0.138        & This Work \\
$\mu^*_\alpha$ (mas yr$^{-1}$)  & $-$7.590\,$\pm$\,0.132        & This Work \\
                                & $-$\,7.597                    & CG20      \\
                                & $-$\,7.595                    & H23       \\
                                & $-$\,7.582                    & Q24 DG    \\
                                & $-$\,7.597                    & Q24 ML    \\
$\mu_\delta$ (mas yr$^{-1}$)    & $+$\,\,2.691\,$\pm$\,0.140    & This Work \\
                                & $+$\,\,2.639                  & CG20      \\
                                & $+$\,2.691                    & H23       \\ 
                                & $+$\,2.689                    & Q24 DG    \\
                                & $+$\,2.686                    & Q24 ML    \\
$\pi$ (mas)                     & 0.744\,$\pm$\,0.035           & This Work \\
                                & 0.719                         & CG20      \\
                                & 0.741                         & H23       \\
                                & 0.742                         & Q24 DG    \\
                                & 0.601                         & Q24 ML    \\
d (kpc)                         & 1.28\,$\pm$\,0.03             & This Work \\    
                                & 1.360                         & CG20      \\
                                & 1.294                         & H23       \\
                                & 1.348                         & Q24 DG    \\
                                & 1.663                         & Q24 ML    \\
(m$-$M) (mag) & 10.53\,$\pm$\,0.05& This Work                               \\    
                                &  10.67                        & CG20      \\
                                & 10.52                         & H23       \\ 
A$_{\rm V}$ (mag)               & 1.11\,$\pm$\,0.02             & This Work \\    
                                & 0.95                          & CG20      \\
                                & 0.95                          & H23       \\
                                & 1.50                          & Q24 DG    \\
                                & 1.40                          & Q24 ML    \\ 
$\log$\,t$_\text{yr}$           & 8.84\,$\pm$\,0.02             & This Work \\    
                                & 8.72                          & CG20      \\
                                & 8.64                          & H23       \\
                                & 8.50                          & Q24 DG    \\
                                & 8.75                          & Q24 ML    \\
$[\text{Fe/H}]$ (dex)           & $-$0.07\,$\pm$\,0.09          & This Work \\
                                & $+$0.06                       & TS23      \\
                                & $+$0.06                       & Q24 DG    \\
                                & $+$0.15                       & Q24 ML    \\
${\text{R}_\text{GC}}$ (kpc)    & \,\,7.95\,$\pm$\,0.16         & This Work \\
M$_\text{turn-off}$ (M$_\odot$) & 2.33\,$\pm$\,0.02             & This Work \\

\hline
\end{tabular}
\end{table}


\subsection{Atmospheric Parameters}

\par For the spectroscopic analysis, we used the \textsc{moog} code \citep[2019 version;][]{sneden1973} as our primary tool. This code allows the spectral line analysis and spectrum synthesis under the assumption of local thermodynamic equilibrium (LTE). Furthermore, we adopted 1D-LTE plane-parallel \textsc{atlas 9} model atmospheres \citep{castelli2004} and used an interpolator to construct more accurate models. Our analysis uses a broad range of absorption lines of Fe\,{\sc i}, Fe\,{\sc ii} and other 21 chemical species analyzed in this work. The line list is available as supplementary material.

To determine the stellar atmospheric parameters, we employed {\sc ares} \citep[Automatic Routine for line Equivalent widths in stellar Spectra;][]{sousa2015},  {\sc iraf} \citep{tody1986}, and the Python package {\sc qoyllur-quipu} \citep[$q^2$;][]{ramirez2014}, which employs 2019 \textsc{moog} version in a semiautomatic way. Initially, equivalent widths (EWs) of Fe\,{\sc i} and Fe\,{\sc ii} lines were manually measured using \texttt{splot} {\sc iraf} routine, and then compared with semiautomatic measures from {\sc ares} version 2. Since we obtained excellent agreement, we used {\sc ares} to obtain EWs of iron lines for our red giants sample. Then, we conducted a spectroscopic study using the results provided by $q^2$.

\par The effective temperature ($T_{\rm eff}$), surface gravity ($\log\,g$), metallicity ([Fe/H]), and microturbulence velocity ($\xi$) were then determined by standard spectroscopic techniques, automatically applied by $q^2$ with manual supervision: $T_{\rm eff}$ was obtained by satisfying the excitation equilibrium; $\xi$ satisfies the condition of no correlation between Fe I abundances and the logarithm of reduced EW ($\log\,{\rm EW}/\lambda$); and $\log\,g$ was determined from the ionization equilibrium. Iron spectral lines were taken from \citet{lambert1996}, which present a list with 165 Fe\,{\sc i} and 24 Fe\,{\sc ii} lines. To select the best quality lines, we then discarded EWs lower than 20 m{\AA} and higher than 150 m{\AA}, and excluded blended lines and those with relative error higher than 10\% of the measurement. In general, lines that are too strong can not be properly adjusted by a gaussian, and lines that are too weak can blend into spectral noise. The best fit parameters of each star are shown on Table \ref{tab:atm}, along with the comparison with other works from the literature, when available.

\par To verify our spectroscopic results, we calculated the temperature parameter using photometric data. The photometric parameters for each star were derived by applying the polynomials provided by \citet{gonzalez2009}. The first step was to obtain the photometric effective temperatures T$_{{\rm eff}}^{\rm (V-K)}$, in which we adopted E(V$-$K)$/$E(B$-$V) $= 2.72$ \citep{mccall2004}. In addition, we constrained the effective temperature using another independent method, the line-depth ratio (LDR) method from \citet{biazzo2007}.

\begin{figure}
\centering
\includegraphics[width=\columnwidth]{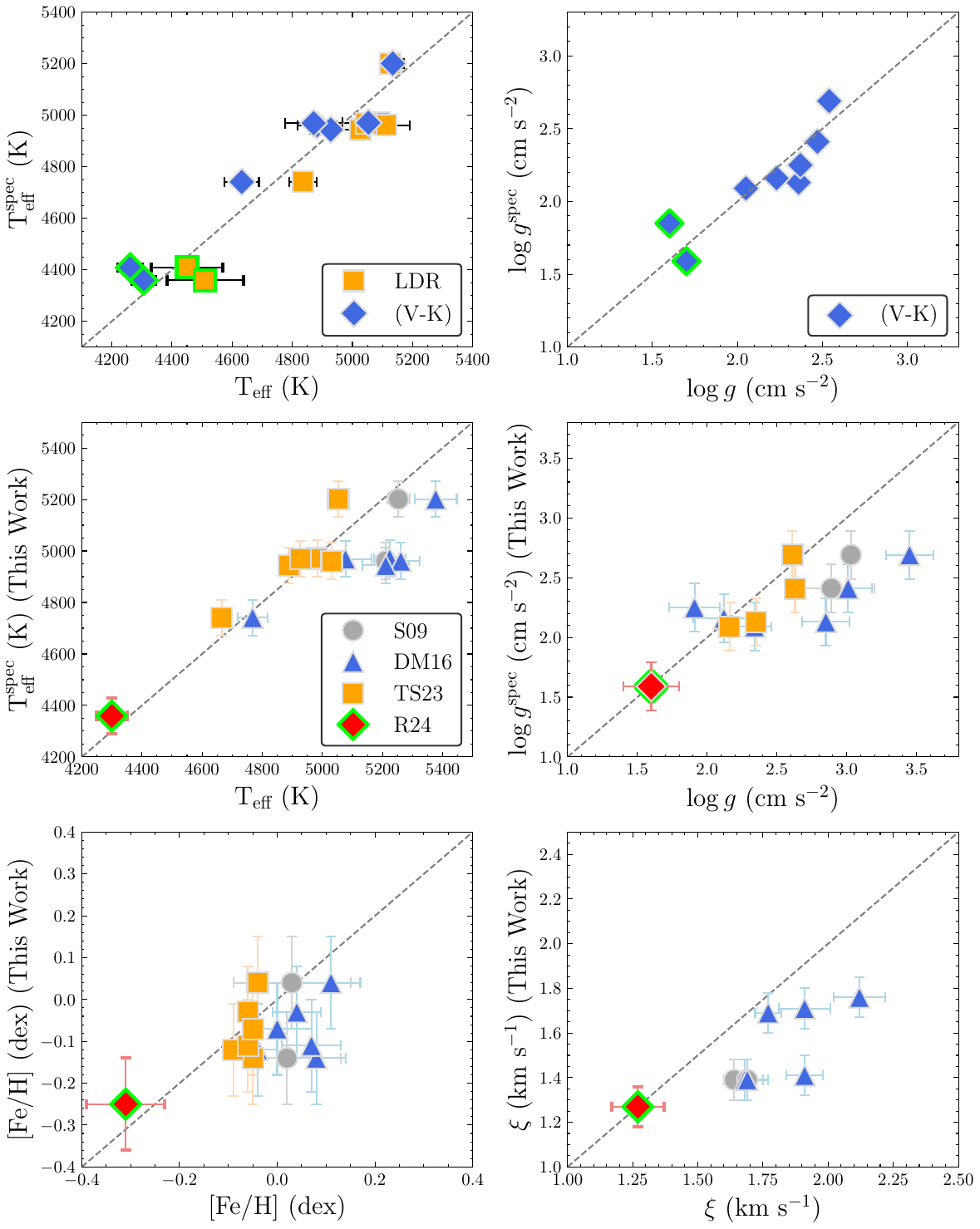}
\caption{Comparison between $T_{{\rm eff}}$ and $\log{g}$ spectroscopic results with those obtained by (V-K) photometry and LDR (top); and comparison between the parameters derived in this work with those from literature (middle and bottom): S09, \citet{santos2009}; DM16, \citet{delgadomena2016}; TS23, \citet{tsantaki2023}; R24, \citet{ramos2024}. The results for the non-member stars (\#028 and \#034) are highlighted in green. The literature comparison plots show only the star \#034, as parameters for the \#028 were not previously reported in the literature.}
\label{fig:params_comparison}
\end{figure}

Also, in order to evaluate our spectroscopic $\log\,g$ results, we estimated the stellar photometric $\log\,g$ using its well-known equation:

\begin{eqnarray*}
\log\,g^{\rm phot}\,=\,\log\,\bigg(\,\frac{M_{{\rm turn-off}}}{M_{\odot}}\,\bigg)\,
+\,0.4\,(K\,-\,A_{\rm K}\,+\,BC_{K})\\
+\,4\,\log\,T_{{\rm eff}}^{\rm (V-K)}\,-\,2\,\log\,d\,({\rm kpc})\,-\,16.5,
\end{eqnarray*}
where $K$, $A_{\rm K}$, and $BC_{\rm K}$ represent the apparent magnitude, interstellar extinction, and bolometric correction in the $K$ band, respectively. The surface gravity, turn-off mass, and effective temperature of the star are denoted by $\log\,g$, $M_{\rm turn-off}$, and $T_{\rm eff}^{\rm (V-K)}$, respectively, while $d$ is the heliocentric distance to the OC. In this expression, we adopted the bolometric correction from \citet{masana2006} and an extinction ratio of $A_K/A_V = 0.114$ \citep{mccall2004}. To estimate the photometric $\log\,g$ uncertainties, we calculated the error propagation including the errors of the parameters measured in this work. In general, we observe a good agreement between our spectroscopic and photometric results for all the stars in our sample, as seen in Figure \ref{fig:params_comparison} (top).

\par In Figure \ref{fig:params_comparison} (middle and bottom), we also present the comparison between our results and those from the literature. We observe a very good agreement with the results from \citet{tsantaki2023} and \citet{ramos2024}. The larger deviations are present when comparing with the results from \citet{delgadomena2016} and some results from \citet{santos2009}, probably caused by methodological differences in these studies.

We underline the fact that only one work in the literature reports atmospheric parameters derived from high-resolution spectroscopy for the star \#034 \citep{ramos2024}, and the parameters for the star \#028 are here presented for the first time. Both are classified as non-members by our astrometric analysis ($p\,<\,0.700$), as discussed in Section \ref{sec:discuss}.

\begin{table*} 
\centering
\caption{Stellar atmospheric parameters obtained from spectroscopy and photometry, and comparison with literature results. Typical uncertainties in $T_{\rm eff}^{\rm spec}$, log\,$g^{\rm spec}$, and $\xi_{t}$ are 70\,K, 0.16\,dex, and 0.06\,km\,s$^{-1}$, respectively. Columns 8 and 10 present, respectively, the number of the spectral lines of Fe\,{\sc i} and Fe\,{\sc ii} considered in the analysis. References: S09, \citet{santos2009}; DM16, \citet{delgadomena2016}; TS23, \citet{tsantaki2023}.} 
\label{tab:atm}
\begin{tabular} {l c c c c c c c c c c c c}
\hline
Star & $T_{\rm eff}^{\rm (V-K)}$ & log\,$g^{\rm phot}$ & $T_{\rm eff}^{\rm LDR}$ & $T_{\rm eff}^{\rm spec}$ & log\,$g^{\rm spec}$ & [Fe\,{\sc i}/H] & \# & [Fe\,{\sc ii}/H] & \# & $\xi_{t}$ & Ref.  \\
 & K & dex & K & K & dex & dex &  & dex &  & km\,s$^{-1}$ \\ \hline
\#005 & 5053\,$\pm$\,86 & 2.36\,$\pm$\,0.05 & 5090\,$\pm$\,19 & 4971 & 2.13 & $-$0.03\,$\pm$\,0.09 & 37 & $-$0.04\,$\pm$\,0.10 & 6 & 1.41 & This Work \\
 & -- & -- & -- &  5224 & 2.85 & $+$0.04\,$\pm$\,0.05 & -- & -- & -- & 1.91 & DM16 \\
 & -- & -- & -- & 4983 & 2.35 & $-$0.06\,$\pm$\,0.02 & -- & -- & -- & -- & TS23 \\
\#028$^\star$ & 4261\,$\pm$\,39 & 1.60\,$\pm$\,0.05 & 4450\,$\pm$\,118 & 4409 & 1.85 & $-$0.26\,$\pm$\,0.10 & 30 & $-$0.27\,$\pm$\,0.07 & 4 & 2.23 & This Work \\
\#034$^\star$ & 4306\,$\pm$\,39 & 1.70\,$\pm$\,0.05 & 4510\,$\pm$\,127 & 4359 & 1.59 & $-$0.25\,$\pm$\,0.11 & 40 & $-$0.26\,$\pm$\,0.12 & 4 & 1.27 & This Work \\
\#053 & 4882\,$\pm$\,65 & 2.47\,$\pm$\,0.05 & 5111\,$\pm$\,80 & 4961 & 2.41 & $-$0.14\,$\pm$\,0.11 & 32 & $-$0.14\,$\pm$\,0.08 & 3 & 1.39 & This Work \\
 & -- & -- & -- & 5211 & 2.89 & $+$0.02\,$\pm$\,0.11 & -- & -- & -- & 1.64 & S09 \\
 & -- & -- &  -- &5261 & 3.01 & $+$0.08\,$\pm$\,0.06 & -- & -- & -- & 1.68 & DM16 \\
 & -- & -- &  -- &5031 & 2.63 & $-$0.05\,$\pm$\,0.02 & -- & -- & -- & -- & TS23 \\
\#087 & 5134\,$\pm$\,36 & 2.54\,$\pm$\,0.04 & 5126\,$\pm$\,29 & 5201 & 2.69 & $+$0.04\,$\pm$\,0.11 & 40 & $+$0.04\,$\pm$\,0.11 & 5 & 1.39 & This Work \\
 & -- & -- &  -- &5253 & 3.03 & $+$0.03\,$\pm$\,0.12 & -- & -- & -- & 1.69 & S09 \\
 & -- & -- &  -- &5377 & 3.45 & $+$0.11\,$\pm$\,0.06 & -- & -- & -- & 1.69 & DM16 \\
 & -- & -- &  -- &5053 & 2.61 & $-$0.04\,$\pm$\,0.02 & -- & -- & -- & -- & TS23 \\
\#121 &4632\,$\pm$\,57 & 2.05\,$\pm$\,0.05 & 4835\,$\pm$\,46 & 4741 & 2.09 & $-$0.12\,$\pm$\,0.10 & 36 & $-$0.11\,$\pm$\,0.11 & 3 & 1.69 & This Work \\
 & -- & -- & -- & 4768 & 2.34 & $-$0.04\,$\pm$\,0.04 & -- & -- & -- & 1.77 & DM16 \\
 & -- & -- & -- & 4665 & 2.16 & $-$0.09\,$\pm$\,0.02 & -- & -- & -- & -- & TS23 \\
\#126 & 4928\,$\pm$\,70 & 2.23\,$\pm$\,0.05 & 5027\,$\pm$\,36& 4944 & 2.16 & $-$0.11\,$\pm$\,0.12 & 42 & $-$0.11\,$\pm$\,0.10 & 4 & 1.76 & This Work \\
 & -- & -- &  -- &5211 & 3.21 & $+$0.07\,$\pm$\,0.06 & -- & -- & -- & 2.12 & DM16 \\
 & -- & -- &  -- &4888 & 2.23 & $-$0.06\,$\pm$\,0.02 & -- & -- & -- & -- & TS23 \\
\#190 & 4871\,$\pm$\,96 & 2.37\,$\pm$\,0.05 & 5046\,$\pm$\,20& 4969 & 2.25 & $-$0.07\,$\pm$\,0.10 & 36 & $-$0.07\,$\pm$\,0.08 & 6 & 1.71 & This Work \\
 & -- & -- &  -- &5077 & 2.82 & $+$0.00\,$\pm$\,0.07 & -- & -- & -- & 1.91 & DM16 \\
 & -- & -- &  -- &4927 & 2.36 & $-$0.05\,$\pm$\,0.02 & -- & -- & -- & -- & TS23 \\
\hline
\end{tabular}
    \begin{tablenotes}
      \small
      \item Note: non-member stars are marked with $\star$.
    \end{tablenotes}
\end{table*}

Errors in the photometric temperature were derived by standard error propagation applied to the polynomials from \citet{gonzalez2009}. Regarding the spectroscopic parameters,  spectroscopic temperature and microturbulence velocity, errors in $T_{\rm eff}^{\rm spec}$ and $\xi_{t}$ were estimated using the slopes found in the aforementioned    methodology, while log\,$g^{\rm spec}$ errors were derived from the standard deviation of the metallicity given by $q^2$. We have found typical uncertainties of $\sigma_{T_{\rm eff}}$\,=\,$\pm$ 70\,K,\, $\sigma_\xi$\,=\,$\pm$\,0.1\,km\,s$^{-1}$, and $\sigma_{\log g}$\,=\,$\pm$\,0.2.

\subsection{Chemical Abundances}

Besides Fe\,{\sc i} and Fe\,{\sc ii}, we investigated the abundances of 14 species using the EW measurement technique via the {\sc iraf} \texttt{splot} routine: Na\,{\sc i}, Mg\,{\sc i}, Al\,{\sc i}, Si\,{\sc i}, Ca\,{\sc i}, Ti\,{\sc i}, Cr\,{\sc i}, Y\,{\sc ii}, Ni\,{\sc i}, Zr\,{\sc i}, La\,{\sc ii}, Ce\,{\sc ii}, Nd\,{\sc ii} and Sm\,{\sc ii}. We adopted the list compiled by \citet{heiter2021} and the lines were selected using the following priority order: i) \textit{gfflag}$=$Y $+$ \textit{synflag}$=$Y; ii) \textit{gfflag}$=$Y $+$ \textit{synflag}$=$U; iii) \textit{gfflag}$=$U $+$ \textit{synflag}$=$Y or U. The parameters \textit{gfflag} and \textit{synflag} were adopted respectively to indicate the relative quality of $\log{\textit{gf}}$ and blending quality, with Y standing for the best quality and U for undecided. We employed specific criteria to select the lines, such as isolation and EW lower than 150\,m\AA. Also, we selected Mg\,{\sc i}, Nd\,{\sc ii} and Sm\,{\sc ii} lines from the line list adopted by \citet{holanda2019, holanda2021}.  Furthermore, we employed spectral synthesis using {\sc moog}'s {\sc synth} routine to determine the chemical abundance of seven species and the $^{12}$C/$^{13}$C ratio: Li\,{\sc i}, C (C$_{2}$), N (CN), [O\,{\sc i}], Sc\,{\sc ii}, Ba\,{\sc ii} and Eu\,{\sc ii}.

Lithium abundances determination was performed by the analysis of the resonance doublet at 6708\,\AA. The line list was constructed using the same data source as \citet{ghezzi2009}. Non-LTE (NLTE) corrections were applied following the grids of \citet{lind2009}, yielding an average correction of $+0.19$ dex. 

\par For oxygen, we performed the spectral synthesis of the 6300{\,\AA} forbidden line. Since the spectra of three stars (\#087, \#126 and \#190) present contamination by telluric lines, for them we assumed the average oxygen abundance of the cluster members.

\par Carbon abundances were determined according \citet{drake2008}, using spectral synthesis at 5635\,{\AA} region of ${\rm A}^3\,\Pi_g$\,$-$\,${\rm X}^3\Pi_u$ molecular band of ${\rm C}_2$(0,1) Swan system. For nitrogen, we obtained the abundances by the set of $^{12}{\rm CN}$ lines in the 8002$-$8006\,{\AA} region of the ${\rm A}^2\,\Pi$\,$-$\,${\rm X}^2\Sigma$ band, using the line list suggested by \citet{carlberg2012}. Furthermore, we determined the $^{12}{\rm C}/^{13}{\rm C}$ isotopic ratio using the $^{13}{\rm CN}$ lines at 8005\,{\AA} region. Final results were obtained iteratively, considering the interdependence between CNO abundances. Sodium abundances were derived using the $\lambda\lambda$6154$-$6160 doublet. Although this feature is known to suffer LTE deviations, corrections from \citet{lind2011} provide negligible deviations for our stars: the average difference between LTE and NLTE was found to be 0.02 dex. Magnesium NLTE abundances were obtained according \citet{osorio2015} and \citet{osorio2016}.

\par Regarding heavier elements, scandium was analyzed by the spectral synthesis of the 5657\,{\AA} line, considering atomic parameters from \citet{lawler2019}. Yttrium abundances were determined using a set of 6 selected medium-EP lines, and LTE deviations were evaluated according \citet{storm2023}. However, Y\,{\sc ii} NLTE corrections are most significant for low-EP spectral lines and very metal-poor stars. For the stars here studied, at solar metallicity, corrections were not greater than $+0.07$ and $+0.03$\,dex for the non-member and member stars, respectively. For barium, considering the atomic data from \citet{mcwilliam1994}, we measured the 5853\,{\AA} line, for which LTE deviations are negligible \citep{korotin2015}. Lanthanum abundances were obtained with the {\sc moog}'s \textit{blends} driver, with atomic parameters from \citet{lawler2001} and \citet{roriz2021}. Europium abundances were derived by the spectral synthesis of the Eu\,{\sc ii} 6656\,{\AA} line and atomic data from \citet{lawler2001}. The adopted atomic parameters and measured EWs are made available as online supplementary material.

\par
\begin{figure}
\centering
\includegraphics[width=\columnwidth]{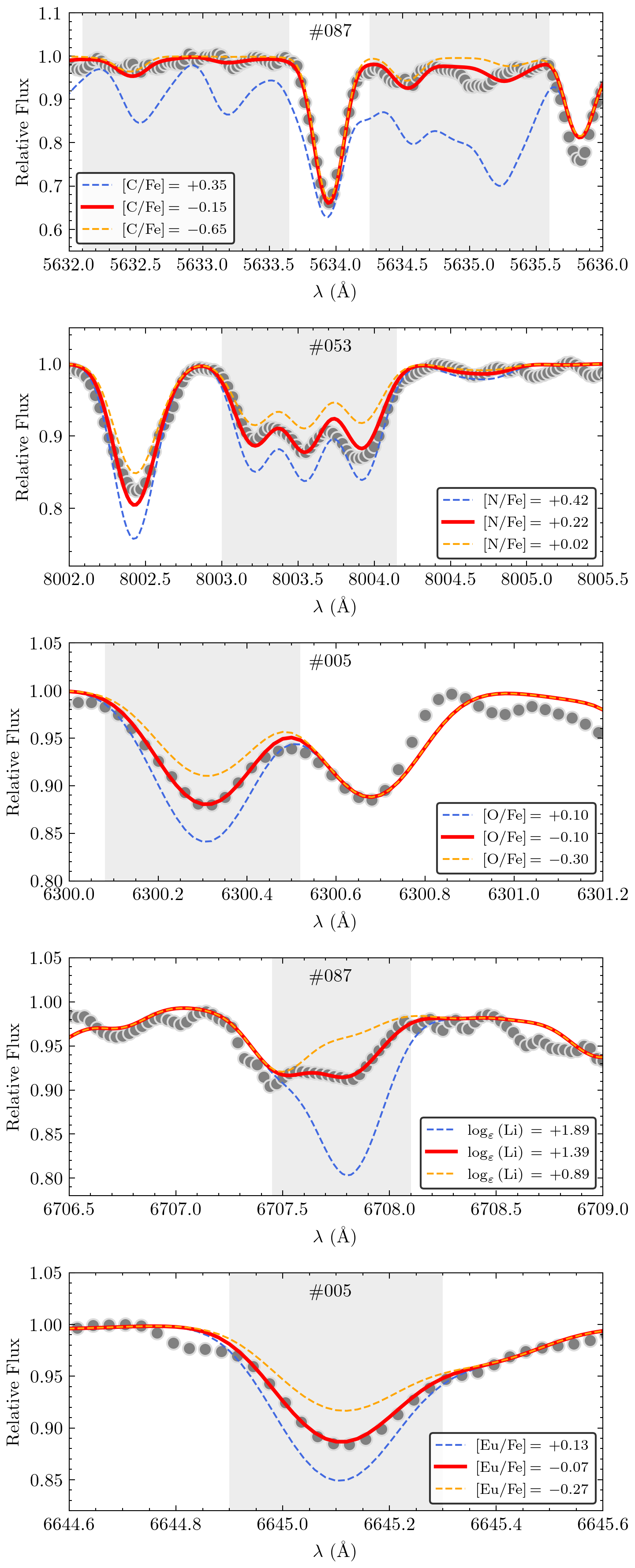}
\caption{Observed data (gray) and different synthetic spectra. The best fit is shown in red. The gray box shows the spectral region considered for the fit.}
\label{fig:ab_synth}
\end{figure}

 Figure \ref{fig:ab_synth} illustrates the spectral synthesis method used in this work, where the theoretical models and the observed data (in gray) are compared. The best fits are shown in red. 

\par The abundances results for light elements are presented in Table \ref{tab:ab_light_el}, which also contains the $^{12}$C/$^{13}$C and the [C/N] ratios. For the heavier elements, the results are shown in Table \ref{tab:ab_el}, together with the mean results for $\alpha$-, iron peak-, \textit{s}- and \textit{r}-process elements. The [X/Fe] abundance ratios were normalized to the solar values according \citet{grevesse2007}. In comparison with the other stars in the sample, we can see how the stars \#028 and \#034 present deviant results for almost all species, which would compromise the mean results for the cluster, significantly raising the standard deviations. For this reason, the cluster mean values are computed excluding both stars. 

\subsection{Uncertainties}

\par We also conducted an analysis to assess the uncertainties in the determination of the chemical abundances and, for this intent, the star \#005 was taken as representative of our sample. Following \citet{holanda2024a}, we estimated atmospheric parameters associated errors. The uncertainty in effective temperature was derived from the error in the slope of the [Fe/H] versus excitation potential relation, while the microturbulence velocity uncertainty was obtained from the slope of [Fe/H] versus reduced equivalent width relation. Surface gravity uncertainty was evaluated by iteratively adjusting $\log{g}$ value until the difference between Fe\,{\sc i} and Fe\,{\sc ii} mean abundances matched the standard mean deviation of [Fe/H]. These uncertainties were then propagated to quantify the final abundances errors. The total uncertainty, $\sigma_{\text{atm}}$, is given as the sum of the squared uncertainties of each atmospheric parameter:
\begin{align*}
    \sigma_{\text{atm}}^2 = & \text{ } \sigma_{T_{\text{eff}},\log{\varepsilon}\text{(X)}}^2 + \sigma_{\log g, \log{\varepsilon}\text{(X)}}^2 + \sigma_{\Delta\log{\varepsilon}\text{(Fe)}, \log{\varepsilon}\text{(X)}}^2 \\ 
    & + \sigma_{\xi_t, \log{\varepsilon}\text{(X)}}^2.
\end{align*}

The final abundance errors are shown in Table \ref{tab:ab_uncert}. In the table, columns 2 to 5 represent the changes in the abundances due to the variation ($\Delta$) of each atmospheric parameter. The adopted variations are $+$70 K, $+$0.16 dex, $+0.09$ dex and $+$0.06 km\,s$^{-1}$ for effective temperature, logarithm of superficial gravity, metallicity and microturbulence velocity, respectively. Considering that the main atmospheric parameters and SNR of spectra are rather similar for the member stars of the cluster, the results for the star \#005 are also suited to be applied to these stars. 

\subsection{Projected Rotational Velocity}

Due to the unknown inclination angle of a star's rotation axis, the projected rotational velocity ($v\,\sin{i}$) provides the best observational proxy for stellar rotation, besides being a valuable indicator for identifying tidal synchronization in close binary systems. Moreover, rapid rotation has often been linked to lithium enrichment. Stars with $v\,\sin{i}\,\geq\,8.0\,$km\,s$^{-1}$ are commonly considered as rapid rotators \citep{drake2002, katime-santrich2013, massarotti2008}, although they are rare, comprising only about 2\% of Galactic red giants \citep{carlberg2011}.

\par In this work, we have estimated the projected rotational velocities using the spectral synthesis technique. We have selected four iron lines sufficiently isolated: $\lambda\lambda$5848, 6151, 6301, and 6302. Measurements with errors greater than 1$\sigma$ were discarded. The macroturbulent velocity was fixed at $3.0$\,km\,s$^{-1}$ \citep{fekel1997} and we adopted FWHM\,$\approx 0.13$\,{\AA}, according to the spectrograph calibration. The member stars provide $\langle v\,\sin{i} \rangle\,=\,5.45\,\pm\,1.48$ \,km\,s$^{-1}$, which agrees with the $v\,\sin{i}$$-$age distribution indicated by \citet{holanda2021}. Our results are shown in Table \ref{tab:vsini}, along with the results obtained by \citet{delgadomena2016}. We can observe that the values found in this work are slightly smaller than the results from the literature, who presented $\langle v\,\sin{i} \rangle\,=\,6.56\,\pm\,1.06$\,km\,s$^{-1}$, considering the same stars.

\par We highlight the results for the stars \#005, \#087 and \#126, which present slightly high projected rotational velocities, that are also observed in the literature. These stars are not reported as binaries, and also their \texttt{ruwe} and RV measurements does not indicate binary companions. Therefore, the presence of chemical anomalies could indicate that their rotational velocities may be due to external sources of angular momenta, such as mergers with companions and engulfment of planets. In \ref{subsec:abundances}, we discuss two possible scenarios that could be unveiled by a chemical tracer.

\begin{table}
\caption{Projected rotational velocities  determined in this work for the red giants in our sample. A comparison with the results from \citet{delgadomena2016}$^{a}$ is also shown, when available}.
\label{tab:vsini}
\begin{tabular} {l c c c c c c c c c c}
\hline
Star & $v\,\sin\,i$ & $v\,\sin\,i^{a}$ \\
 & km\,s$^{-1}$ & km\,s$^{-1}$ \\
\hline                                                              
\#005 & 6.18\,$\pm$\,0.74 & 7.24 \\
\#028$^\star$ & 2.59\,$\pm$\,0.86 & -- \\
\#034$^\star$ & 2.91\,$\pm$\,0.92 & -- \\
\#053 & 5.00\,$\pm$\,0.49 & 6.45 \\
\#087 & 6.70\,$\pm$\,0.82 & 7.61 \\
\#121 & 2.76\,$\pm$\,0.85 & 4.54 \\
\#126 & 6.65\,$\pm$\,1.18 & 7.47 \\
\#190 & 5.39\,$\pm$\,0.69 & 6.10 \\
\hline
\end{tabular}
    \begin{tablenotes}
      \small
      \item Note: non-member stars are marked with $\star$.
    \end{tablenotes}
\end{table}


\section{Discussion} \label{sec:discuss}

\subsection{Membership and Cluster Parameters} \label{sec:membership_results}

The astrometric analysis conducted in this work resulted in membership probabilities for 4498 stars in the field of the open cluster IC\,2714, 898 of which were classified as members (p\,$\geq\,0.700$). Using an artificial neural network (ANN) applied to Gaia\,DR2 data, \citet{Cantat2020} found 888 members for this cluster. \citet{hunt2023} employed the Hierarchical Density-Based Spatial Clustering of Applications with Noise \citep[HDBSCAN,][]{campello2023} to analyze the membership of thousands of clusters in the Gaia\,DR3 dataset, and presented 1285 members for IC\,2714. More recently, this cluster's membership was assessed by \citet{qiu2024} from two methods: the algorithm DG, a combination of the DBSCAN \citep{ester1996} with the Gaussian mixture model \citep[GMM,][]{figueiredo2002}; and the maximum likelihood. The authors found 1097 and 1419 members from the DG algorithm and the maximum likelihood method, respectively. Therefore, our membership results better align with those from the ANN and DG methods from literature. Despite the differences between the number of cluster members, in general our main cluster parameters align well with those from the literature, as can be seen in Table \ref{tab:general}. Particularly, between the methods employed by \citet{qiu2024}, our results present a better agreement with those derived from the DG method, with the exception of the extinction and age parameters. The deviations between our work and the literature are probably due to differences in the employed methodologies.

\par In particular, regarding our sample of eight red giants, we found that six of them can be classified as members using the astrometric probability criteria. The exceptions are the stars \#028 and \#034. The former presents a rather low membership probability (of about 0.09), while for the latter it was not possible to assign a reliable probability since its discrepant proper motions does not meet the adopted criteria of the quality filters. Furthermore, both stars present parallaxes that are incompatible with the cluster population at their magnitude range (see Figure \ref{fig:4plots}). This is interesting, considering that both stars were classified as cluster members by \citet{mermilliod2008} due to their radial velocities (RV). Besides, using data from Gaia\,DR2, \citet{Cantat2020} classified \#028 as member with probability p $= 1$. 

\par In addition to membership probabilities for each star, the Table \ref{tab:general} presents, in column 10, the notes assigned by \citet{mermilliod2008}. All the six member stars are marked as single (S) by the authors. On the other hand, the stars \#028 and \#034 are marked as possible spectroscopic binaries (SB). Since all stars present \texttt{ruwe} $< 1.4$ (column 6), it is possible that the Gaia observations could not detect their companions and thus is treating them as single stars. The variation observed in the radial velocity results of the star \#028 (see columns 7, 8 and 9 in Table \ref{tab:general}) could strengthen its spectroscopic binary classification, which could be later addressed with new radial velocities measurements. Furthermore, in Figure \ref{fig:4plots}, we can see how \#028 and \#034 (in green) present in general discrepant astrometric results in comparison with the others giants of our sample (in red); the star \#034 is found outside the limits of the proper motions distribution plot. Thus, our membership analysis indicate that stars \#028 and \#034 are non-members. Besides being included in our abundances results (Tables \ref{tab:ab_light_el} and \ref{tab:ab_el}), these stars were disregarded in the evaluation of the average results of the cluster and in the isochrone fitting. As we discuss in \ref{subsec:abundances}, our chemical abundances results corroborate the classification of these stars as non-members.

From the membership analysis, we derived metallicity, distance, $\log$\,t$_\text{yr}$ and A$_{\rm V}$ by isochrone fitting, as described in Section \ref{sec:meth}. Assuming $8.34 \pm 0.16$ kpc for the solar Galactocentric distance \citep{reid2014}, we obtained R$_\text{GC} = 7.95 \pm 0.16$ for IC\,2714. The derived parameters are presented in Table \ref{tab:general}, in comparison with the results from \citet{Cantat2020}, \citet{tsantaki2023} and \citet{hunt2023} when available. We can see how our results agree with the literature, considering the uncertainties. In particular, the six member stars (\#005, \#053, \#087, \#121, \#126 and \#190) provide an average iron abundance of $\langle \text{[Fe/H]} \rangle \,= -$0.07\,$\pm$\,0.09. Considering the uncertainties, this result agrees with the metallicity derived via isochrone fitting ([Fe/H] = $-0.18 \pm 0.05$). Furthermore, our result approaches that from \citet{tsantaki2023}, of $\langle \text{[Fe/H]} \rangle \,= +$0.06\,$\pm$\,0.02.

\begin{table*}
\begin{threeparttable}
\caption{Abundance ratios [X/Fe] of the light elements and carbon isotopic ratio of the red giants of IC\,2714.}
\label{tab:ab_light_el}
\begin{tabular} {c c c c c c c c c c }
\hline 
Ratio & $\#$005 & $\#$053 & $\#$087 & $\#$121 & $\#$126 & $\#$190 & Mean & $\#$028$^\star$ & $\#$034$^\star$ \\
\hline
$\log\,\varepsilon$(Li)             & $+$\,0.96              & $+$\,1.07         & $+$\,1.39 & $+$\,0.30 & $+$\,1.08   & $+$\,0.96   & $+$0.96\,$\pm$\,0.33 & $+$\,0.13 & $+$\,0.62 \\
$\log\,\varepsilon$(Li)$_{\rm NLTE}$& $+$\,1.15              & $+$\,1.25         & $+$\,1.54 & $+$\,0.54 & $+$\,1.27   & $+$\,1.15   & $+$1.15\,$\pm$\,0.30 & $+$\,0.44 & $+$\,0.95 \\
{[C/Fe]}\,(C$_{2}$)                 &$-$\,0.11               & $-$\,0.13         & $-$\,0.15 & $-$\,0.26 & $-$\,0.10   & $-$\,0.15   & $-$0.15\,$\pm$\,0.05 & $+$\,0.02 & $-$\,0.29 \\
{[N/Fe]}\,(CN)                      &$+$\,0.22               & $+$\,0.27         & $+$\,0.30 & $+$\,0.46 & $+$\,0.32   & $+$\,0.37   & $+$0.32\,$\pm$\,0.08 & $+$\,0.35 & $+$\,0.09 \\
{[O/Fe]}                            &$-$\,0.17               & $-$\,0.08         & $-$\,0.22$^{**}$ & $+$\,0.01 & $-$\,0.06$^{**}$   & $-$\,0.01$^{**}$   & $-$0.09\,$\pm$\,0.08 & $+$\,0.18 & $+$\,0.06 \\
{[C/N]}            &$-$\,0.33               & $-$\,0.40         & $-$\,0.45 & $-$\,0.72 & $-$\,0.42   & $-$\,0.52   & $-$0.48\,$\pm$\,0.12 & $-$\,0.33 & $-$\,0.38 \\
$^{12}$C/$^{13}$C                   & 14                     & 18                & 17       & 16        & 19          & 13          & 16\,$\pm$\,2         & 16        & 11        \\
\hline 
\end{tabular}
    \begin{tablenotes}
      \small
      \item Note 1: non-member stars are marked with $\star$.
      \item Note 2: Stars with mean oxygen abundance ($\log\,\varepsilon$(O) $= +8.53$) assumed are marked with ${**}$.
    \end{tablenotes}
  \end{threeparttable}
\end{table*}

\begin{table*}
\begin{threeparttable}
\caption{Abundance ratios [X/Fe] of elements Na to Eu of the red giants of IC\,2714.}
\label{tab:ab_el}
\begin{tabular} {c c c c c c c c c c}
\hline 
Ratio &  $\#$005 &  $\#$053 &  $\#$087 &  $\#$121 &  $\#$126 &  $\#$190 &  Mean &  $\#$028$^\star$ &  $\#$034$^\star$\\
\hline
{[Na\,{\sc i}/Fe]}              & $+$\,0.18 & $+$\,0.15 & $+$\,0.09 &  $-$\,0.08 & $+$\,0.27 & $+$\,0.10  &  $+$\,0.12\,$\pm$\,0.11 & $+$\,0.03 & $-$\,0.09 \\
{[Na\,{\sc i}/Fe]$_{\rm NLTE}$} & $+$\,0.17 & $+$\,0.13 & $+$\,0.04 &  $-$\,0.06 & $+$\,0.26 & $+$\,0.08  &  $+$\,0.10\,$\pm$\,0.10 & $+$\,0.09 & $+$\,0.00 \\
{[Mg\,{\sc i}/Fe]}              & $+$\,0.05 & $-$\,0.03 & $-$\,0.13 &  $+$\,0.08 & $+$\,0.12 & $-$\,0.04  &  $+$\,0.01\,$\pm$\,0.08 & $+$\,0.07 & $+$\,0.03 \\
{[Al\,{\sc i}/Fe]}              & $+$\,0.25 & $+$\,0.31 & $-$\,0.03 &  $+$\,0.12 & $+$\,0.07 & $+$\,0.13  &  $+$\,0.14\,$\pm$\,0.11 & $+$\,0.14 & $+$\,0.20 \\      
{[Si\,{\sc i}/Fe]}              & $+$\,0.01 & $+$\,0.16 & $+$\,0.21 &  $+$\,0.07 & $+$\,0.20 & $+$\,0.14  &  $+$\,0.16\,$\pm$\,0.09 & $+$\,0.31 & $+$\,0.21 \\
{[Ca\,{\sc i}/Fe]}              & $-$\,0.02 & $+$\,0.14 & $+$\,0.16 &  $+$\,0.05 & $-$\,0.06 & $+$\,0.02  &  $+$0.05\,$\pm$\,0.08   & $+$\,0.04 & $+$\,0.16 \\
{[Sc\,{\sc i}/Fe]}              & $-$\,0.20 & $-$\,0.08 & $-$\,0.08 &  $-$\,0.07 & $+$\,0.02 & $-$\,0.04  &  $-$0.08\,$\pm$\,0.07   & $+$\,0.05 & $-$\,0.02 \\
{[Ti\,{\sc i}/Fe]}              & $-$\,0.08 & $+$\,0.05 & $+$\,0.08 &  $+$\,0.11 & $-$\,0.07 & $+$\,0.00  &  $+$\,0.02\,$\pm$\,0.07 & $+$\,0.19 & $+$\,0.21 \\
{[Cr\,{\sc i}/Fe]}              & $+$\,0.00 & $+$\,0.04 & $+$\,0.06 &  $+$\,0.09 & $-$\,0.13 & $-$\,0.15  &  $-$\,0.01\,$\pm$\,0.10 & $+$\,0.19 & $-$\,0.01 \\
{[Ni\,{\sc i}/Fe]}              & $+$\,0.06 & $-$\,0.06 & $-$\,0.03 &  $+$\,0.00 & $-$\,0.05 & $-$\,0.08  &  $-$\,0.03\,$\pm$\,0.05 & $-$\,0.03 & $+$\,0.19 \\
{[Y\,{\sc ii}/Fe]}              & $+$\,0.04 & $+$\,0.02 & $+$\,0.05 &  $+$\,0.07 & $-$\,0.08 & $-$\,0.11  &  $+$\,0.00\,$\pm$\,0.07      & $+$\,0.14 & $-$\,0.15 \\
{[Zr\,{\sc i}/Fe]}              & $+$\,0.03 & $-$\,0.05 & $+$\,0.12 &  $+$\,0.15 & $+$\,0.02 & $-$\,0.07  &  $+$\,0.07\,$\pm$\,0.12 & $+$\,0.34 & $+$\,0.02 \\
{[Ba\,{\sc ii}/Fe]}             & $+$\,0.13 & $+$\,0.10 & $+$\,0.20 &  $-$\,0.14 & $-$\,0.12 & $-$\,0.12  &  $+$\,0.01\,$\pm$\,0.14 & $-$\,0.41 & $+$\,0.09 \\
{[La\,{\sc ii}/Fe]}             & $+$\,0.11 & $+$\,0.19 & $+$\,0.14 &  $+$\,0.19 & $+$\,0.17 & $+$\,0.12  &  $+$\,0.15\,$\pm$\,0.03 & $+$\,0.35 & $+$\,0.17 \\
{[Ce\,{\sc ii}/Fe]}	            & $-$\,0.09 & $+$\,0.08 & $+$\,0.12 &  $-$\,0.07 & $-$\,0.17 & $-$\,0.28  &  $-$\,0.07\,$\pm$\,0.14 & $-$\,0.07 & $+$\,0.25 \\
{[Nd\,{\sc ii}/Fe]}             & $+$\,0.30 & $+$\,0.12 & $+$\,0.04 &  $+$\,0.15 & $+$\,0.07 & $-$\,0.03  &  $+$\,0.11\,$\pm$\,0.10 & $+$\,0.38 & $+$\,0.21 \\
{[Sm\,{\sc ii}/Fe]}             & $+$\,0.05 & $+$\,0.10 & $-$\,0.16 &  $+$\,0.04 & $-$\,0.13 & $-$\,0.16  &  $-$\,0.04\,$\pm$\,0.11      & $+$\,0.06 & $+$\,0.20 \\
{[Eu\,{\sc ii}/Fe]}             & $-$\,0.03 & $-$\,0.06 & $+$\,0.00 &  $-$\,0.02 & $+$\,0.06 & $+$\,0.10  &  $+$\,0.01\,$\pm$\,0.05 & $+$\,0.32 & $+$\,0.02 \\
{[odd/Fe]}                      & $+$\,0.21 & $+$\,0.22 & $+$\,0.01 &  $+$\,0.03 & $+$\,0.17 & $+$\,0.11  &  $+$\,0.12\,$\pm$\,0.08 & $+$\,0.12 & $+$\,0.11 \\
{[$\alpha$/Fe]}                 & $-$\,0.01 & $+$\,0.08 & $+$\,0.08 &  $+$\,0.08 & $+$\,0.04 & $+$\,0.03  &  $+$\,0.05\,$\pm$\,0.03 & $+$\,0.15 & $+$\,0.15 \\
{[peak/Fe]}                     & $-$\,0.05 & $-$\,0.03 & $-$\,0.02 &  $+$\,0.01 & $-$\,0.07 & $-$\,0.09  &  $-$\,0.04\,$\pm$\,0.03 & $+$\,0.07 & $+$\,0.06 \\
{[$s$/Fe]}                      & $+$\,0.09 & $+$\,0.08 & $+$\,0.11 &  $+$\,0.06 & $-$\,0.02 & $-$\,0.08  &  $+$\,0.04\,$\pm$\,0.07 & $+$\,0.12 & $+$\,0.11 \\
{[$r$/Fe]}                      & $+$\,0.01 & $+$\,0.02 & $-$\,0.08 &  $+$\,0.03 & $-$\,0.03 & $-$\,0.03  &  $+$\,0.01\,$\pm$\,0.04 & $+$\,0.19 & $+$\,0.15 \\
\hline 
\end{tabular}
    \begin{tablenotes}
      \small
      \item Note: non-member stars are marked with $\star$.
    \end{tablenotes}
  \end{threeparttable}
\end{table*}

\begin{table}
\centering
\caption{Abundance uncertainties for the representative star \#005.}
\label{tab:ab_uncert}
\begin{tabular}{l c c c c c}
\hline
El. & {\scriptsize $\Delta T_{\rm eff}$} & {\scriptsize $\Delta \log\,g$} & {\scriptsize $\Delta$ [Fe/H]} & {\scriptsize $\Delta \xi_{t}$} & $\sigma_{\rm atm}$ \\ 
  & {\scriptsize $+70$\,K} & {\scriptsize $+0.16$\,dex} & {\scriptsize $+0.09$\,dex} & {\scriptsize $+0.06$\,km\,s$^{-1}$} & \\ \hline
C\,{\scriptsize (C$_{2}$)} & $+$\,0.05 & $+$\,0.00 & $-$\,0.01 & $+$\,0.00 & 0.05 \\
N\,{\scriptsize (CN)} & $+$\,0.00 & $-$\,0.09 & $-$\,0.06 & $-$\,0.12 & 0.16 \\
O\,{\sc i} & $-$\,0.04 & $+$\,0.08 & $+$\,0.00 & $-$\,0.06 & 0.11 \\
Li\,{\sc i} & $+$\,0.18 &  $+$\,0.10 & $+$\,0.09 & $+$\,0.10 & 0.25 \\
Na\,{\sc i} & $+$\,0.09 & $-$\,0.04 & $+$\,0.01 & $+$\,0.00 & 0.10 \\
Mg\,{\sc i}& $+$\,0.03 & $-$\,0.01  & $+$\,0.00 & $-$\,0.02 & 0.04 \\
Al\,{\sc i}& $+$\,0.05 & $-$\,0.01  & $+$\,0.00 & $-$\,0.01 & 0.05 \\
Si\,{\sc i}& $+$\,0.01 & $+$\,0.02  & $+$\,0.01 & $-$\,0.01 & 0.03 \\
Ca\,{\sc i} & $+$\,0.07 & $-$\,0.02 & $-$\,0.01 & $-$\,0.03 & 0.08 \\
Sc\,{\sc ii} & $+$\,0.00 & $+$\,0.09 & $+$\,0.00 & $-$\,0.02 & 0.09 \\
Ti\,{\sc i} & $+$\,0.09 & $-$\,0.01 & $-$\,0.01 & $-$\,0.01 & 0.09 \\
Cr\,{\sc i} & $+$\,0.07 & $-$\,0.01 & $+$\,0.00 & $-$\,0.02 & 0.07 \\
Ni\,{\sc i} & $+$\,0.06 & $+$\,0.02 & $+$\,0.01 & $-$\,0.03 & 0.07 \\
Y\,{\sc ii} & $+$\,0.00 & $+$\,0.07 & $+$\,0.03 & $-$\,0.04 & 0.09 \\
Ba\,{\sc ii} & $+$\,0.03 & $+$\,0.05 & $+$\,0.01 & $-$\,0.08 & 0.10 \\
Zr\,{\sc i} & $+$\,0.12 & $-$\,0.01 & $-$\,0.01 & $+$\,0.00 & 0.12 \\
La\,{\sc ii} & $+$\,0.01 & $+$\,0.07 & $+$\,0.03 & $-$\,0.02 & 0.07 \\
Ce\,{\sc ii} & $+$\,0.01 & $+$\,0.07 & $+$\,0.03 & $-$\,0.01 & 0.08 \\
Nd\,{\sc ii} & $+$\,0.01 & $+$\,0.07 & $+$\,0.03 & $-$\,0.02 & 0.08 \\
Sm\,{\sc ii} & $+$\,0.02 & $+$\,0.07 & $+$\,0.03 & $-$\,0.02 & 0.08 \\
Eu\,{\sc ii} & $+$\,0.00 & $+$\,0.07 & $+$\,0.01 & $-$\,0.01 & 0.07 \\ \hline
\end{tabular}

\end{table}

\subsection{Abundances} \label{subsec:abundances}

\subsubsection{C, N, Na and isotopic ratio}

\begin{figure}
    \centering
    \includegraphics[width=\columnwidth]{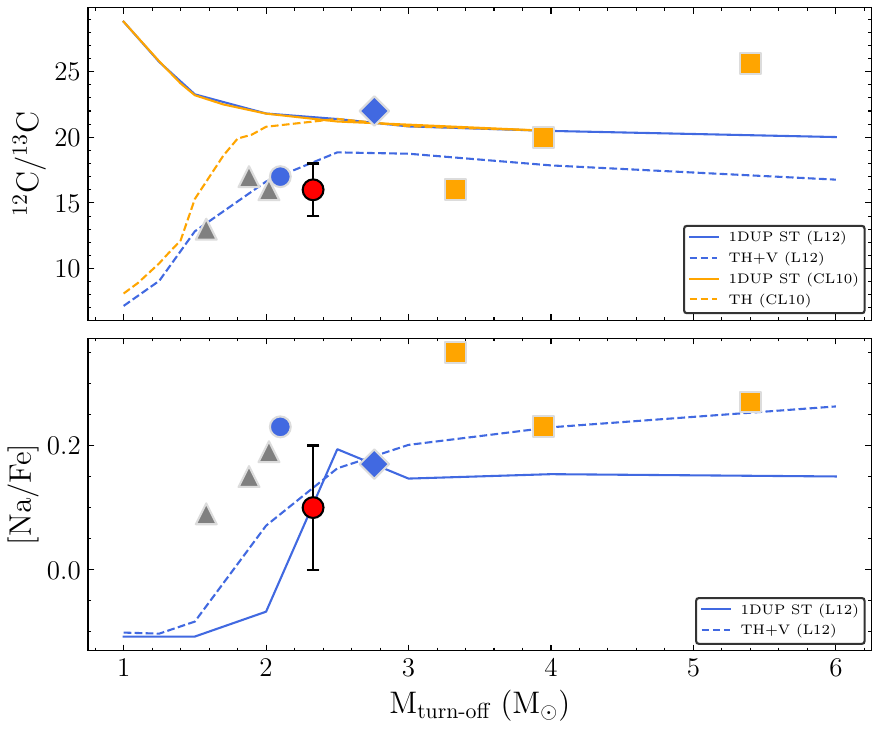}
    \caption{$^{12}$C/$^{13}$C isotopic ratio (top) and [Na/Fe] ratio (bottom) in terms of the turn-off mass of giant stars in OCs. Our results for IC\,2714 are shown in red, and the mean [Na/Fe] ratio was corrected by NLTE effects. The results for OCs previously studied by our team are also shown: \citet{penasuarez2018}, in gray triangles; \citet{dasilveira2018}, in blue diamonds; \citet{martinez2020}, in blue circles; and \citet{holanda2019, holanda2021, holanda2022}, in orange squares. Solid lines represent the predicted abundances at the first dredge-up from solar metallicity evolutionary models from CL10 \citep[][orange]{charbonnel2010} and L12 \citep[][blue]{lagarde2012}. Dashed lines shows the prediction models for thermohaline and rotation-induced mixing (TH+V) from L12 (in blue), thermohaline extra-mixing from CL10 (orange), and TH+V from L12 (in blue), respectively.}
    \label{fig:ab_mixing}
\end{figure}

We expect from the FDU a significant increase of the stellar $^{14}$N abundance, a $\sim$30\,\% decrease of $^{12}$C abundance, and an increase of $^{13}$C abundance \citep{karakas2014}. However, further extra-mixing events could deviate the observable stellar abundances from the classical predictions. The prediction models of \citet{charbonnel2010} and \citet{lagarde2012} indicate that the effect of the FDU and thermohaline mixing on sodium abundance is very sensitive to the stellar mass: we expect low enrichment in low-mass stars, while intermediate-mass stars must present higher sodium abundances in both models, as can be seen in Figure \ref{fig:ab_mixing} (bottom). Considering only the member stars, the average [Na/Fe] ratio of the cluster agrees with both first dredge-up (1DUP\,ST) and thermohaline mixing with rotation-induced mixing (TH+V) models from \citet{lagarde2012}, considering the uncertainties. Results from OCs studied by our team are also shown, including data from \citet{penasuarez2018}, \citet{dasilveira2018}, \citet{martinez2020} and \citet{holanda2019, holanda2021, holanda2022}.  

\par The $^{12}$C/$^{13}$C ratio can be used as a tracer of nucleosynthesis and stellar evolution in low- and intermediate-mass stars. From the FDU, classical models predict a ratio of $^{12}$C/$^{13}$C $\sim 22$ for a $\sim 2.3 $ M$_\odot$ star, as can be seen in Figure \ref{fig:ab_mixing} (top). However, our average result of 16\,$\pm$\,2 for the IC\,2714 members is even below the TH+V model from \citet{lagarde2012}, indicating that the stars experienced both the thermohaline and rotation-induced mixing during their evolution. Due to the quite large masses of IC\,2714 stars, the rotation-induced mixing was dominating. This could indicate that the analyzed giants are RC stars, in the core-helium-burning phase.

\begin{figure*}
    \centering
    \includegraphics[scale=0.7]{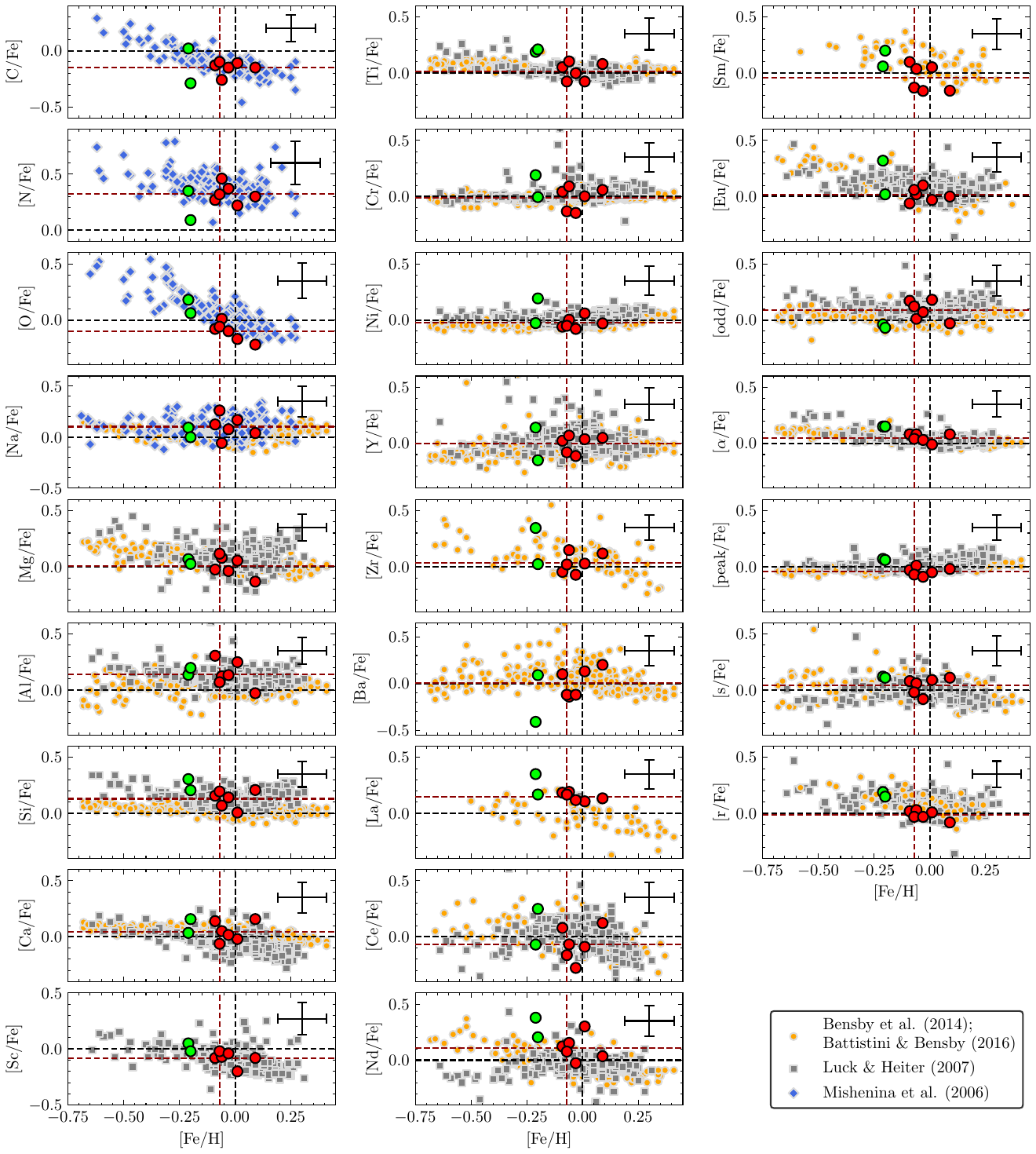}
    \caption{Abundance ratios [X/Fe] versus [Fe/H] for elements from carbon to europium. [$\alpha$/Fe], [odd/Fe], [peak/Fe], [s/Fe], and [r/Fe] represent the averages for $\alpha$-elements (Si, Ca, Ti, Mg), odd-Z elements (Na, Al), iron-peak elements (Sc, Cr, Ni), elements predominantly produced by the \textit{s}-process (Y, Zr, Ba, La, Ce, Nd), and elements predominantly produced by the \textit{r}-process (Sm, Eu), respectively. The results for the stars classified as members are shown in red, while those for the non-member stars are shown in green. In blue diamonds results by \citet{mishenina2006}; in gray squares by \citet{luck2007}; and in orange circles by \citet{bensby2014} and \citet{battistini2016}. The intersection of the black dashed lines indicate solar values, while the intersection of the red lines indicate the average abundances of the members.}
    \label{fig:ab_ratios}
\end{figure*}

In Figure \ref{fig:ab_ratios}, we compare abundances of carbon, nitrogen and sodium in member (red) and non-member stars (green) with solar neighborhood G$-$K giants from \citet{luck2007} (gray squares) and red giants from \citet{mishenina2006} (blue diamonds). For sodium, our results are also compared with solar neighborhood dwarfs from \citet{bensby2014} and \citet{battistini2016}. Among our sample, the non-member stars \#028 and \#034 display notably distinct behavior. The former shows the highest [C/Fe] ratio, while the latter presents the lowest. In terms of [N/Fe], \#028 appears consistent with the cluster members, but the star \#034 is once again an outlier showing the lowest result. When placed in the context of literature values, the star \#028 lies within the typical range given the uncertainties, whereas \#034 shows significantly discrepant carbon and nitrogen abundances. These chemical results further support the non-member classification of the stars \#028 and \#034, reinforcing the astrometric evidence earlier presented in Section \ref{subsec:abundances}.

\subsubsection{Lithium}

The analysis of lithium abundances yields a powerful diagnostics for a range of stellar and Galactic astrophysical processes, including stellar structure and evolution, internal mixing mechanisms and age estimation, and thus they can be used to trace star formation history in OCs \citep[e.g][]{randich2021}. Also, as mentioned in Section \ref{sec:intro}, lithium abundances can provide valuable insights in the current knowledge of chemically peculiar stars.

\par From the previous high-resolution spectroscopy studies of IC\,2714 that had the lithium abundance as a target, \citet{delgadomena2016} reported three lithium-rich ($\log\,\varepsilon$(Li)\,$>$\,1.50\,dex) stars that are also present in our sample. The authors studied Li-rich giants distributed across 12 OCs, and presented lithium abundances for 67 stars, including six stars here analyzed.  For the stars \#053, \#087 and \#126, they presented NLTE lithium abundances of 1.56, 1.63 and 1.60\,dex, respectively. To explain the lithium enrichment in these stars, the authors proposed that these hottest objects are least evolved and thus are at an early phase of FDU, having not yet experienced the expected lithium dilution. In this way, these stars could be incorrectly cataloged as Li-rich giants. More recently, \citet{tsantaki2023} analyzed the same data and reported revised lithium abundances of 1.29, 1.36 and 1.27\,dex for the stars \#053, \#087 and \#126, respectively. Our measurements are in closer agreement with these updated values. As suggested by \citet{carlberg2016}, the higher lithium results in \citet{delgadomena2016} likely resulted from overestimated effective temperatures, as clearly illustrated in Figure \ref{fig:params_comparison}. This establish their results as outliers relative to our findings and those from other studies.

In this work, our membership and chemical analysis allowed a thorough investigation regarding the true nature of these stars. As discussed in \ref{sec:membership_results}, with our membership analysis we were able to build a well-defined main sequence for IC\,2714, and also to conduct a reliable isochrone fitting. These analyses are fundamental to correctly determine the evolutionary stages of the stars. Furthermore, with the study of a larger set of chemical species, we are able to further investigate evolutionary stages based on the expected nucleosynthesis. The indicative that our analyzed stars are in the RC phase, as suggested by its $^{12}$C/$^{13}$C ratios, is also corroborated by its positions in the CMD (Figure \ref{fig:4plots}).

Based on spectroscopic and asteroseismic observations, several works from the literature indicated a predominance of Li-rich giants in the RC phase \citep[e.g.][]{deepak2021, yan2021, singh2019, singh2021, casey2019}. Particularly, \citet{kumar2020} proposed the presence of a systematic lithium production in low-mass stars between the tip of the RGB and the RC. To evaluate this scenario, \citet{magrini2021b} discussed the lithium abundances of RC stars in open clusters using the sixth internal data release of the Gaia-ESO Survey \citep[IDR6,][]{gilmore2012, randich2013}. They discuss two possible mechanisms to explain the lithium enrichment in RC stars: the mixing induced by the first He flash, and the mixing induced by neutrino magnetic moment. However, both models may not be totally necessary to explain the lithium abundances in the younger clusters. Interestingly, the data analyzed by the authors corroborate with a possible Li enrichment during the RC phase. Also, they suggest that there might be a lithium enrichment event between the RGB and RC phases that results in stars that would not be typically classified as Li-rich, being this scenario corroborated by the $^{12}$C/$^{13}$C ratios \citep{magrini2021b}. Since the carbon isotopic ratios of the here analyzed IC\,2714 giants agree with the presence of an extra-mixing, this could suggest lithium enrichment even for stars whose lithium abundances are below the traditional limit.

Considering the traditional limit, we found one lithium-rich giant within the six IC\,2714 members: the star \#087 with $\log\,\varepsilon$(Li)$_{\rm NLTE}$\,$=$\,1.54\,dex. In the light of some of the current proposed scenarios in the literature, the relatively high projected rotational velocity of this star ($v\,\sin{i}\,=\,$ 6.70\,km\,s$^{-1}$) together with its high lithium abundance could corroborate a possible accretion scenario such as proposed by \citet{siess1999a, siess1999b}. Also, its lithium abundance is found within the expected upper limit for this model, \citep[2.2\,dex,][]{aguilera2016}. As mentioned in Section \ref{sec:intro}, Be and B abundances could help to confirm this model, but the spectral signatures of these elements fall outside {\sc feros} coverage, at around 3100{ \AA}, and therefore were not analyzed in this work. A merger scenario such as proposed by \citet{zhang2020} may also be considered to explain the high lithium abundance of the star \#087. However, since rotational velocities are not considered in this model, the high $v\,\sin{i}$ of this lithium-rich giant could not be used to explain its nature in this case. On the other hand, this merger scenario could also be corroborated by the possible RC nature of this star.

It is worth highlighting the possible relation between lithium enrichment and infrared excess, which could be used to identify Li-rich giants \citep[e.g.][]{delareza1996, delareza1997, kumar2015}. Remarkably, when comparing with non-rich giants, infrared excesses are at least twice as common in Li-rich K giants, and thus the mechanism that is responsible for this enrichment could be associated to dust production \citep{rebull2015}. In this sense, infrared photometry could also be a valuable asset to constrain and elucidate the nature of lithium-rich stars.

\subsubsection{O, Al, \texorpdfstring{$\alpha$}{alpha}- and iron-peak elements}

For oxygen, all the stars in our sample agree with the expected trend within the uncertainties. However, we observe that the non-members exhibit elevated [O/Fe] relative to the cluster average. Considering the uncertainties, we observe a good agreement between the aluminum abundances of the members and the literature data. We found a mean abundance of $\langle \text{[Al/Fe]} \rangle \,= +$0.14\,$\pm$\,0.11, which indicates a chemical homogeneity for this species within the member stars of IC\,2714.

The $\alpha-$elements, such as Mg, Si, Ca, and Ti, are produced in short-time scale events, mainly by type II, Ib and Ic supernovae \citep{kobayashi2020}. On the other hand, iron-peak elements such as Sc, Cr and Ni, are predominantly produced in significantly longer-time scales, in type Ia supernova explosions.

Regarding a long-lasting issue in astrophysics that is the determination of stellar ages, $\alpha-$elements have been widely explored in the context of the so called chemical clocks. In particular, as discussed in the end of this section, the combination between these elements and those from \textit{s}-process (see below) has shown promise in establishing relations between chemical abundances and stellar ages in solar neighborhood \citep[e.g.][]{spina2016, delgadomena2019, tautvaisiene2021}, and were also extended to different regions of the Galactic disc \citep[e.g.][]{viscasillas2022}. Thus, the study of $\alpha-$elements abundances can provide valuable results to current studies in the context of chemical clocks related to Galactic chemical evolution.

\par Our results for the cluster members present an average $\alpha-$elements abundance near solar, of $\langle \text{[$\alpha$/Fe]} \rangle \,= +$0.11\,$\pm$\,0.03. We can observe how the non-member stars \#028 and \#034 present slightly higher abundances when compared to the cluster average, but also agree with the literature trend considering uncertainties. This behavior is also seen when considering iron-peak elements: the non-member giants show increased abundances when compared to the cluster average. We found an average result close to solar, with $\langle \text{[peak/Fe]} \rangle \,= -$0.04\,$\pm$\,0.03. Given that the $\alpha$- and iron-peak elements presented the smallest abundances dispersions, these elements might be good candidates for establishing a cluster chemical signature (e.g. chemical membership).

\subsubsection{s- and \textit{r}-process elements}

Heavier elements produced via neutron-capture processes fall into two categories: the \textit{s}-process and the \textit{r}-process. The \textit{s}-process elements are produced by slow neutron capture primarily during TP-AGB phase in low- and intermediate-mass stars \citep{bisterzo2016, karakas2014}. On the other hand, although its exact astrophysical site remains debated, the \textit{r}-process is associated with rapid neutron capture mainly in more energetic events, including Type II supernovae, neutron stars mergers and neutron star-black hole mergers \citep[e.g.][]{kobayashi2020, woosley1994}.

Similar to $\alpha-$elements, \textit{s}-process elements are also used in the context of chemical clocks. Studies have shown a trend of increasing \textit{s}-process enrichment with decreasing stellar age, motivating the use of [s/$\alpha$] ratios, such as [Y/Mg], as empirical age indicators for both solar twins and open clusters across the Galactic disc \citep[e.g.][]{nissen2010,nissen2016,nissen2020,dorazi2009, dorazi2022, maiorca2011, maiorca2012, casamiquela2021}.

\par In this work, we analyzed the abundances of the \textit{s}-process-dominated elements Y, Zr, Ba, La, Ce and Nd, and \textit{r}-process elements Sm and Eu. Literature data indicate significant \textit{s}-process enrichment in young OCs with ages up to 150$-$200 Myr \citep[e.g.][]{dorazi2009, dorazi2022, maiorca2011, maiorca2012}, which can be traced by [Ba/Fe] ratios as high as 0.5$-$0.6 dex. However, our estimated cluster age of $\sim$690 Myr aligns with our results, which show no such enrichment. The cluster members yield an average barium abundance of $\langle \text{[Ba/Fe]} \rangle \,= +0.01\,\pm$\,0.14, with the star \#087 displaying a modest enhancement of [Ba/Fe] $= +0.19$.

Although the standard deviations are relatively high for some species, particularly for [Ba/Fe], the overall average abundance for \textit{s}-process elements is $\langle \text{[s/Fe]} \rangle \,= +0.04\,\pm$\,0.07, which is consistent with solar values and with the literature results. For the \textit{r}-process elements, our results for Sm and Eu are also in good agreement with the expected from the literature. We report a mean abundance of $\langle \text{[r/Fe]} \rangle \,= +0.01\,\pm$\,0.04\,dex, further supporting the chemical homogeneity among the red giants population of IC\,2714.

\subsubsection{Galactocentric chemical gradients}

As discussed, OCs are found in a wide range of ages and metallicities, being distributed throughout the whole Galactic disc. They provide ideal sites to study several astrophysical phenomena and can be used as chemical tracers to address certain gaps in the current knowledge of stellar formation and evolution, in addition to Galaxy chemical evolution models \citep{netopil2016, spina2021, magrini2023}. Figure \ref{fig:feh_rgc} shows the metallicity gradient with OC data from \citet{magrini2023}, including our results (in red). The [Fe/H] derived from IC\,2714 members is well aligned with the observed trend for young clusters ($t\,<\,$ 1 Gyr), which is also consistent with its estimated age. The IC\,2714 is found within the $R_\text{GC}$ range of 6$-$10 kpc, for which we observe the unusual trend of younger clusters being less metal-rich than older ones. Besides, the gradient is flatter for the younger clusters within this same range. These trends are corroborated by several works from literature \citep[e.g.][]{magrini2009, randich2022, spina2016}, and are strengthen by Gaia\,DR3 data, as pointed out by \citet{recio-blanco2023}. In general, our results reinforce the trends identified in \citet{magrini2023}, corroborating the discussion presented therein.

\begin{figure}
    \centering
    \includegraphics[width=\columnwidth]{figures/fig6.pdf}
    \caption{Metallicity [Fe/H] as a function of the Galactocentric distance R$_\text{GC}$ for IC\,2714 (red circle) and other OCs in the Galactic disc, color-coded by their age $t$. The orange line shows the weighted linear regression considering only the younger clusters (t $< 1$ Gyr), and its uncertainties are given by the orange area. OCs data and regression coefficients were taken from \citet{magrini2023}. The dashed lines indicate the solar values.}
    \label{fig:feh_rgc}
\end{figure}

\subsubsection{Chemical clocks}

The determination of stellar ages is fundamental to the understanding of Galactic formation and evolution. However, one can not directly measure these ages, and this determination can prove to be one of the most difficult task in astrophysics \citep[e.g.][]{randich2018, soderblom2014}. For this intent, a direct comparison between models and observational data obtained by isochrone fitting is often employed, but this technique is not always able to provide reliable ages for field stars. Therefore, several works search for observational parameters that could act as stellar age tracers and are related with the Galactic chemical evolution. Some chemical ratios have been widely adopted to this intent in literature, such as [C/N], [Y/Mg], [Y/Al], [Y/Si], [Y/Ca] and [Y/Ti] \citep[e.g.][]{storm2023,spoo2022,feuillet2018, ho2017, ness2016, feltzing2017, slumstrup2017, spina2018, viscasillas2022, casali19, katime-santrich2022}.

\par
\begin{figure}
    \centering
    \includegraphics[width=\columnwidth]{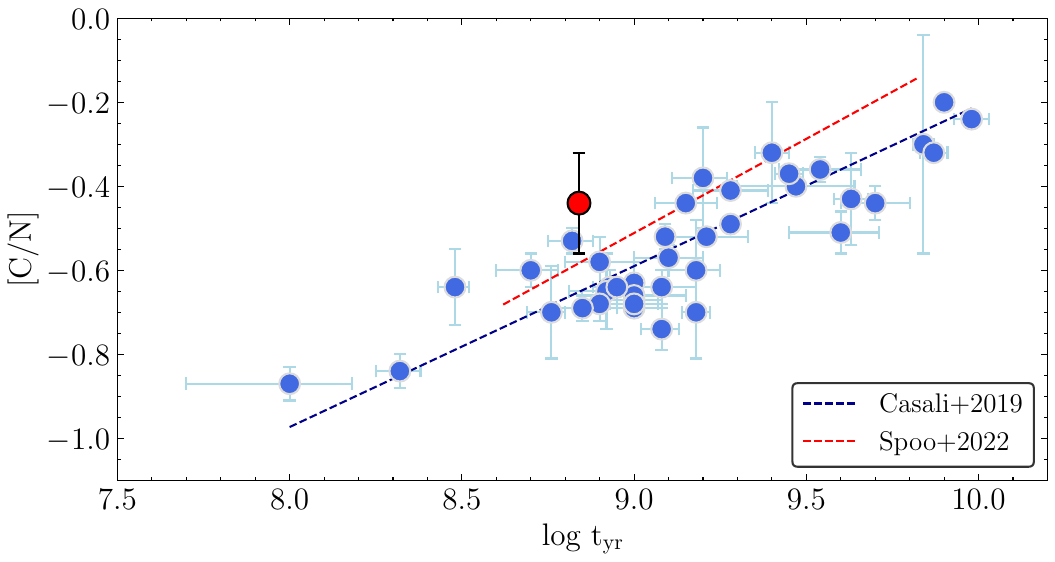}
    \caption{[C/N] ratio as function of $\log{t_\text{yr}}$ for open clusters analyzed by \citet{casali19}. The blue dashed line shows the linear weighted least-squares fit for the data, while the red dashed line shows the calibration from \citet{spoo2022}. Our results for IC\,2714 are shown in red.}
    \label{fig:cn_logage}
\end{figure}

 We expect mass-dependent photospheric changes in C and N abundances due the FDU and, following the discussions in \citet{casali19}, the [C/N] ratio may be employed as stellar age tracer. In this sense, Figure \ref{fig:cn_logage} shows the empirical calibration (blue dashed line) between [C/N] ratio and stellar age presented by \citet{casali19}, combining Gaia-ESO and APOGEE Data Release 12 (DR12) data (blue circles). Our average [C/N] result for IC\,2714 is shown in red. Also, we compare this calibration with that of \citet{spoo2022}, derived from APOGEE DR17 data (red dashed line). Regarding our results, we observe a better agreement with the calibration from DR17. Also, estimating the stellar [C/N] ratio using both calibrations, we obtain, respectively, $-0.65$ and $-0.58$ for \citet{casali19} and \citet{spoo2022} relations, the latter yielding a better agreement with our average cluster ratio of [C/N] $= -0.48\,\pm\,0.12$.

\par On the other hand, the combination of \textit{s}-process elements with $\alpha$-process elements is able to maximize their correlation with stellar age \citep[e.g.][]{spina2016, delgadomena2019, tautvaisiene2021}. In this sense, \citet{katime-santrich2022} studied the [Y/Mg, Al, Si, Ca, Ti] ratios as potential chemical clocks trough a line-by-line spectroscopic analysis of 50 red giants in seven OCs. The authors obtained linear regression and correlation coefficients for both dwarf and giant stars. Using its coefficients, we obtained a good agreement between our results and those from literature. The exception is the [Y/Mg] ratio, for which the literature reports a wide scattering, mainly observed in younger OCs \citep[e.g.][]{blanco-cuaresma2015, reddy2019}. Further studies are required to investigate this trend.

\section{Conclusions}\label{sec:conc}

We conducted a comprehensive analysis of eight red giants reported as members of the OC IC\,2714. Firstly, updating previously results from the literature, we used Gaia\,DR3 data to provide a list of 898 cluster members between the stars with G\,$< 18$ mag in the field of the cluster. We present the cluster's mean proper motions ($\mu^*_\alpha$, $\mu_\delta$) = ($-$7.590, $+$2.691) mas yr$^{-1}$ and parallax $\pi = 0.744$ mas yr$^{-1}$. Using isochrone fitting, we derived age $\log$\,t$_\text{yr} = $ 8.84, distance d$ = 1.28\pm0.03$ kpc, and  interstellar extinction A$_\text{V} = 1.11 \pm 0.02$ mag. All the stars of our sample were previously classified as cluster members by \citet{mermilliod2008}. Nevertheless, we here classified the stars \#028 and \#034 as non-members based on its astrometric probabilities, and they were thus removed from our subsequent analysis. Both show discrepant proper motions and parallaxes and are classified as possible spectroscopic binaries in \citet{mermilliod2008}, but our results weaken this classification for the star \#034. This star was also reported as IC\,2714 member in a more recent work \citep{Cantat2020}. However, the non-member classification of the stars \#028 and \#034 is corroborated by our extensive chemical analysis, which reveals the general discrepant features of these stars in comparison with the cluster members.

\par We derived stellar atmospheric parameters via high-resolution spectroscopy, and they are here presented for the star \#028 for the first time. Our spectroscopic parameters were validated using photometric data and line-depth ratio \citep[LDR, ][]{biazzo2007}, two independent approaches that reinforce our results. Also, from the derived projected rotational velocities, we report three stars with a relatively high rotation, the stars \#005, \#087 and \#126. The results for the stars \#005 and \#126 do not indicate lithium enrichment. However, the star \#087 deserves special attention, since presents a high lithium abundance ($\log_\varepsilon\text{(Li)}_\text{NLTE} = 1.54$ dex). Two possible scenarios could be employed to address the lithium enrichment of the star \#087: a planet or substellar companion accretion event, which would be favored by its high rotational velocity \citep{siess1999a, siess1999b}; and a merger event between a RGB and a He-white dwarf, which would be favored by the possible RC nature of this star \citep{zhang2020}. Further analysis is required to assess the validity of these two models in explaining the lithium abundance observed in this star.

\par Regarding chemical abundances, we here presented the most thorough analysis of IC\,2714 to date, analyzing the stellar abundances of 23 species: C, N, O\,{\sc i}, Li\,{\sc i}, Na\,{\sc i}, Mg\,{\sc i}, Al\,{\sc i}, Si\,{\sc i}, Ca\,{\sc i}, Sc\,{\sc ii}, Ti\,{\sc i}, Cr\,{\sc i}, Fe\,{\sc i}, Fe\,{\sc ii}, Ni\,{\sc i}, Y\,{\sc ii}, Ba\,{\sc ii}, Zr\,{\sc i}, La\,{\sc ii}, Ce\,{\sc ii}, Nd\,{\sc ii}, Sm\,{\sc ii} and Eu\,{\sc ii}. Also, we here presented $^{12}$C/$^{13}$C ratios for the first time for the studied stars. Our chemical results are validated by comparison with extensive literature data, with which the six member stars of our sample present excellent general agreement. We show that the \citet{lagarde2012} model that combine thermohaline and rotation-induced mixing fits well the average IC\,2714 $^{12}$C/$^{13}$C ratio, while both TH+V and first dredge-up models could explain its average sodium abundance. Our results suggest that the analyzed stars may be core-helium-burning red clump stars, and thus, even stars with A(Li)\,$< 1.5$\,dex could have undergone a lithium enrichment event \citep{magrini2021b}.

Finally, we verified that our average chemical results agree with the galactocentric chemical gradients trends identified by \citet{magrini2023}, corroborating the discussion presented therein. Regarding [C/N] ratio, our results are better fitted by the calibration from \citet{spoo2022}. In summary, our work agrees with the literature trends in the context of chemical gradients in the Galaxy, and may offer contributions in constraining chemical abundance vs. age calibrations. 


\begin{acknowledgments}

This study was financed in part by the Coordenação de Aperfeiçoamento de Pessoal de Nível Superior – Brasil (CAPES) – Finance Code 001, proc. 88887.140791/2025-00. NH acknowledges a fellowship of the Fundação de Amparo à Pesquisa do Estado do Rio de Janeiro - FAPERJ, Rio de Janeiro, Brazil, for grant E-26/200.097/2025. F. Maia acknowledges financial support from Conselho Nacional de Desenvolvimento Cientifico e Tecnologico – CNPq (proc. 404482/2021-0) and from Fundação de Amparo a Pesquisa do Estado do Rio de Janeiro - FAPERJ (proc. E- 26/201.386/2022 and E-26/211.475/2021). BPLF acknowledges financial support from Conselho Nacional de Desenvolvimento Científico e Tecnológico (CNPq, Brazil; procs. 140642/2021-8 and 314718/2025-7) and Coordenação de Aperfeiçoamento de Pessoal de Nível Superior (CAPES, Brazil; Finance Code 001; proc. 88887.935756/2024-00). WJBC acknowledges the support from CNPq–BRICS 440142/2022-9, FAPEMIG APQ 02493–22, and FNDCT/FINEP/REF 0180/22. S.D. acknowledges CNPq/MCTI for grant 306859/2022-0 and FAPERJ for grant 210.688/2024.

This study is based on observations conducted with the 2.2\,m MPG/ESO telescope at the European Southern Observatory (ESO) in La Silla, Chile. These observations were made possible through an agreement between Observatório Nacional and ESO.

This work used data from the European Space Agency (ESA) mission {\it Gaia} (\url{https://www.cosmos.esa.int/gaia}), processed by the {\it Gaia} Data Processing and Analysis Consortium (DPAC, \url{https://www.cosmos.esa.int/web/gaia/dpac/consortium}). Funding for the DPAC has been provided by national institutions, in particular, the institutions participating in the {\it Gaia} Multilateral Agreement.

\end{acknowledgments}


\bibliography{bibliography}{}
\bibliographystyle{aasjournal}

\end{document}